\documentstyle[11pt,axodraw]{article}

\begin{document}

\begin{center}
{\LARGE Massless and massive one-loop three-point functions in negative
dimensional approach}

\vspace{1cm}

{\large A T Suzuki}$^{1,a}${\large , E S Santos}$^{1,b}${\large , A
G M Schmidt}$^{1,2,c}$

\vspace{.5cm}

$^{1}$Instituto de F\'{\i}sica Te\'{o}rica

Universidade Estadual Paulista

Rua Pamplona 145

01405-900 - S\~{a}o Paulo, S.P.

Brazil

\vspace{.5cm}

$^{2}$Universidade Federal do Paran\'{a}

Departamento de F\'{\i}sica

Caixa Postal 19044

Curitiba PR 81531-990

Brazil

\vspace{1cm}

\begin{minipage}{12.5cm}
\centerline{\bf Abstract}

{In this article we present the complete massless and massive one-loop
triangle diagram results using the negative dimensional integration
method (NDIM). We consider the following cases: massless internal
fields; one massive, two massive with the same mass $m$ and three
equal masses for the virtual particles. Our results are given in terms
of hypergeometric and hypergeometric-type functions of external
momenta (and masses for the massive cases) where the propagators in
the Feynman integrals are raised to arbitrary exponents and the
dimension of the space-time $D$.  Our approach reproduces the known
results as well as other solutions as yet unknown in the literature.
These new solutions occur naturally in the context of NDIM revealing a
promising technique to solve Feynman integrals in quantum field
theories.  }
\end{minipage}
\end{center}

\bigskip Keywords: negative dimensional integration, one-loop triangle
diagram

PACS nbs.: 02.90+p, 12.38.Bx 

\bigskip

$^{a}$E-mail:suzuki@ift.unesp.br

$^{b}$E-mail:esdras@ift.unesp.br

$^{c}$E-mail:schmidt@fisica.ufpr.br

\newpage

\section{Introduction}
$ $

The study of scattering amplitudes in Quantum Electrodynamics (QED),
Quantum
 Chromodynamics (QCD), and the Standard Model (SM) for
electroweak
 interactions, as well as the renormalization group, the
asymptotic freedom,
 and other properties of perturbative quantum
field theories, have required
 each time the computation of complex
Feynman integrals. Therefore, the
 development of refined
mathematical methods and approaches to deal with them
 have been
researched and applied to various cases with varied successes.
 Some
of them are: dimensional regularization \cite{hoof/Veltman,
giambiagi},
 use of Mellin-Barnes representation for hypergeometric
function \cite{davydy}
 , method of Gegenbauer polynomials
\cite{terrano}, integration by parts \cite
 {tkachov} and several
others \cite{smirnov}-\cite{fleischer}.

Another integration method that has been developed in recent years
and
 applied with great success is the negative dimensional
integration method \
 (NDIM) \cite{hallyday}. This method uses the
analytic continuation of
 dimension $D$ into negative values. One
advantage from NDIM is that the
 complexities of performing
$D$-dimensional integrals are transferred to a
 resolution of systems
of linear algebraic equations. The NDIM approach was
 proven to be
satisfactory when applied to the calculation of various types
 of
Feynman diagrams at one- and two-loop levels as well as in
noncovariant
 gauges such as the light-cone and Coulomb ones \cite
{suzuki/alexandre1,suzuki/alexandre2,suzuki/alexandre3,suzuki/alexandre4}.

In this paper we use NDIM to obtain the complete set of solutions to
one-loop triangle diagram, that occur in some processes such as
interactions
 between Z particles, gluons, etc.; diagrams which
become more and more
 significant in the precision measurements
leading to the checking of
 electroweak standard model, and search
for the Higgs intermediate boson for
 example. Some cases from this
vertex graph type were investigated with other
 approaches
\cite{davydy}, \cite{davydy1}-\cite{oldenborgh}. On the other
 hand,
with NDIM technique we are able to obtain all the possible different
solutions to one-loop triangle diagram in different kinematical
regions of
 interest according to the convergence region
considered. These solutions are
 expressed in terms of
hypergeometric-type functions.

This work is organized as follows. In the first Section we illustrate
the
 use of NDIM to solve one-loop massless triangle and point out
its solutions
 in Appendix A. Massive cases with two null masses; one
null and two equal
 masses; and three equal masses, are treated in
Section {\bf 2} and the
 solutions presented in Appendices B, C and D
respectively. Wherever possible
 results are then compared to known
ones in the pertinent literature (either
 in specific kinematical
region or particular cases with specific values for
 the exponents of
propagators set to minus one). Finally in the last Section
 we
discuss the main results of the paper and present our concluding
remarks.

\section{One-loop massless triangle}

$ $

In this section, we present the result for the one-loop massless triangle
diagram (see Figure above) with two independent external momenta, evaluated
according to the procedure of NDIM.

\begin{figure}[tbp]
\begin{center}
\vspace{25mm} 
\begin{picture}(200,120)(0,-20)
\ArrowLine(50,0)(0,0)
\ArrowLine(80,52)(50,0)
\ArrowLine(80,52)(80,102)
\ArrowLine(110,0)(80,52)
\ArrowLine(110,0)(50,0)
\ArrowLine(160,0)(110,0)
\Text(25,10)[c]{$p$}
\Text(80,10)[c]{$k$}
\Text(135,10)[c]{$q$}
\Text(45,26)[c]{$p-k$}
\Text(100,77)[c]{$q-p$}
\Text(115,26)[c]{$q-k$}
\Text(80,-20)[c]{Figure}
\end{picture}
\end{center}
\end{figure}

The most general form for the one-loop Feynman integral associated to this
diagram is given by 
\begin{equation}
\int \frac{d^{D}k}{ [k^{2}-m_{1}^{2}]^{a^{\prime}}[(k-p)^{2}-m_{2}^{2}]^{b^{
\prime}}[(k-q)^{2}-m_{3}^{2}]^{c^{\prime}}}\;.  \label{a1}
\end{equation}
where $D>0$ and $a^{\prime}, b^{\prime}, c^{\prime}>0$.

In the context of NDIM we take the corresponding integral, namely, 
\begin{eqnarray}
J & = & J(a,b,c,p,q,m_1,m_2,m_3)  \nonumber \\
& = & \int
d^Dk\;[k^{2}-m_{1}^{2}]^{a}[(k-p)^{2}-m_{2}^{2}]^{b}[(k-q)^{2}-m_{3}^{2}]^{c}
\label{a11}
\end{eqnarray}
with $a,b,c>0$, and $D<0$. The physically interesting result emerges after
analytic continuation into positive dimensionality and negative exponents
for $a,b,c$.

We begin by taking the special case where all the internal field masses are
set to zero, that is, we take $m_{1}=m_{2}=m_{3}=0.$ The starting point is
the evaluation of the corresponding Gaussian integral 
\begin{eqnarray}
I & = & I(\alpha ,\beta ,\gamma ,p,q)  \nonumber \\
&=&\int d^{D}k\exp \{-\alpha k^{2}-\beta (k-p)^{2}-\gamma (k-q)^{2}\}
\label{a2} \\
&=&\left[ \frac{\pi }{\alpha +\beta +\gamma }\right] ^{D/2}\exp \left\{ 
\frac{-\beta \gamma (p-q)^{2}-\alpha \beta p^{2}-\alpha \gamma q^{2}}{\alpha
+\beta +\gamma }\right\}  \label{a3}
\end{eqnarray}
where, after the expansion in $\alpha ,\beta $ and $\gamma $ powers, we get 
\begin{eqnarray}
I &=&\pi ^{D/2}{\sum_{j_{1},...,j_{6}=0}^\infty}(-1)^{j_{1}+j_{2}+j_{3}}
\Gamma (1-j_{1}-j_{2}-j_{3}-D/2)  \nonumber \\
&&\times \frac{\alpha ^{j_{1}+j_{2}+j_{4}}}{j_{4}!}\frac{\beta
^{j_{1}+j_{3}+j_{5}}}{j_{5}!}\frac{\gamma ^{j_{2}+j_{3}+j_{6}}}{j_{6}!}\frac{
(p^{2})^{j_{1}}}{j_{1}!}\frac{(q^{2})^{j_{2}}}{j_{2}!}\frac{(r^{2})^{j_{3}}}{
j_{3}!}.  \label{a4}
\end{eqnarray}
Since we use a multinomial expansion the sum indices above are constrained
by 
\[
D/2=-\sum_{n=1}^{n=6}j_n. 
\]
On the other hand, expanding the exponential of (\ref{a2}), we have

\begin{equation}
I={\sum_{a,b,c=0}^\infty}(-1)^{a+b+c}\frac{\alpha ^{a}\beta ^{b}\gamma ^{c}}{
a!b!c!}J(a,b,c,p,q,m_{1}=m_{2}=m_{3}=0).  \label{b2}
\end{equation}
where, comparing the expressions (\ref{a4}) and (\ref{b2}), by its $\alpha
,\beta $ and $\gamma $ powers, we obtain a general relation for the integral 
$J=J(a,b,c,p,q),$

\begin{eqnarray}
J &=&\pi ^{D/2}(-1)^{-a-b-c}\Gamma (1+a)\Gamma (1+b)\Gamma (1+c)  \nonumber
\\
&\times& {\sum_{j_{1},...,j_{6}=0}^\infty }(-1)^{j_{1}+j_{2}+j_{3}}\frac{
\Gamma (1-j_{1}-j_{2}-j_{3}-D/2)!}{j_{4}!j_{5}!j_{6}!}\frac{(p^{2})^{j_{1}}}{
j_{1}!}\frac{(q^{2})^{j_{2}}}{j_{2}!}\frac{(r^{2})^{j_{3}}}{j_{3}!}, 
\nonumber \\
&&  \label{b3}
\end{eqnarray}
where for convenience we introduced $r=q-p$. From the comparison of powers
we obtain additional three constraint equations besides the one already
defined above coming from the multinomial expansion. Then, all constraints
are 
\begin{eqnarray}
a &=&j_{1}+j_{2}+j_{4},  \label{b4} \\
b &=&j_{1}+j_{3}+j_{5},  \label{b5} \\
c &=&j_{2}+j_{3}+j_{6},  \label{b6} \\
\frac{D}{2} &=&-j_{1}-j_{2}-j_{3}-j_{4}-j_{5}-j_{6},  \label{b7}
\end{eqnarray}
Therefore, there are four constraint equations with six variables. These
form a system of linear equations that can only be solved if we leave two
free indices and the result will be given as a double series. There are $
C_4^6=15$ distinct ways in which we can choose these two free indices. From
these 15 ways there are 12 with non-trivial solutions, since for 3 of the
resulting systems of linear equations, the determinant is zero. These 12
solutions are grouped into three sets, each one with four solutions
according to the kinematical configuration of the variables defined by the
external momenta. Performing the analytic continuation to $D>0$ and $a,b,c<0$
we get the three sets of solutions for the relevant Feynman integral (see
Appendix A, Massless cases)

\begin{center}
\begin{tabular}{|c|c|}
\hline
Set & Solutions \\ \hline\hline
$1 $ & $J_1+J_2+J_3+J_4 $ \\ \hline
$2 $ & $J_5+J_6+J_7+J_8 $ \\ \hline
$3 $ & $J_9+J_{10}+J_{11}+J_{12} $ \\ \hline
\end{tabular}
\end{center}
according to their variables, where $J_{n}$, $n=1,2,...,12$ are written in
terms of Apell hypergeometric functions $F_{4}.$ The first set of solutions
above is in accordance with the result calculated in \cite{davydy}. The other two
new solutions represent other regions of kinematical configuration not
explored explicitly on the text books but with same physical importance. 
These two other hypergeometric series representations for the scalar
integral in question can be easily obtained from set 1 by interchanging two external legs
and the associated exponents of propagators, e.g., $p\leftrightarrow q$ and $
b\leftrightarrow c$. However, the negative dimensional approach can generate
hypergeometric series which \underline{cannot} be obtained through the
symmetry of the diagram. An illustration of this occurs in the scalar 1-loop box integral \cite
{suzuki/alexandre1} for the photon-photon scattering where two of such solutions
(which are not related by symmetry but through direct analytic continuation)
are, 
\[
F_3(...|x,y),\quad {\rm and }\quad H_2(....|x,-y^{-1}). 
\]
where $F_3$ and $H_2$ are the usual hypergeometric functions of two variables.

In the next cases, some hypergeometric series can also be obtained by
symmetry considerations when one of the solutions is known (of course,
no such symmetric solutions can ever be generated if one has nothing
to start with! ...). Upon this given solution we can carry out the
needed interchanges in momenta and exponents of propagators in order
to arrive at the other solutions. We emphasize, however, that besides
these solutions which are connected by symmetry, there are completely
new solutions which cannot be related by such means and they were not
obtained previously by any other method in spite of the fact that they
have the same physical importance. This is the strength of NDIM
technique where we obtain {\em simultaneously} all the solutions:
those which are related simply by symmetry of the diagram and also
those which are related by analytic continuation.

\subsection{One-loop triangle, one mass case}

$ $

Now we analyze a second case that occurs when there is only one massive
propagator within the Feynman integral represented by (\ref{a1}) with $
m_{1}=m$ and $m_{2}=m_{3}=0$. Again, we used negative dimensional integral
method to obtain twenty three non trivial solutions with sixteen sets of
independent solutions for the integral, given in terms of hypergeometric
functions of three variables, shown in Appendix B (One massive propagator
case). The sixteen independent solutions are

\vspace{.25cm}

\begin{center}
\begin{tabular}{|c|c||c|c||c|c||c|c|}
\hline
Set & Solutions & Set & Solutions & Set & Solutions & Set & Solutions \\ 
\hline\hline
$1$ & $J_1+J_2$ & $5$ & $J_7+J_8$ & $9$ & $J_{12}$ & $13$ & $J_{18}$ \\ 
\hline
$2$ & $J_3$ & $6$ & $J_{9}$ & $10$ & $J_{13}+J_{14}$ & $14$ & $
J_{19}+J_{20}$ \\ \hline
$3$ & $J_4$ & $7$ & $J_{10}$ & $11$ & $J_{15}+J_{16}$ & $15$ & $
J_{21}+J_{22} $ \\ \hline
$4$ & $J_5+J_6$ & $8$ & $J_{11}$ & $12$ & $J_{17}$ & $16$ & $J_{23}$ \\ 
\hline
\end{tabular}
\end{center}

\vspace{.25cm}

The first solution, namely, $J_{1}+J_{2}$, is in accordance with the result
calculated in \cite{davydy}. Note that there are fifteen other novel results
defined in different kinematical regions not computed anywhere else.

\subsection{One-loop triangle, two equal masses case}
$ $

We consider also the case of two massive propagators with equal mass, where $
m_{1}=0$ and $m_{2}=m_{3}=m$. We have obtained thirty two independent
solutions expressed through the hypergeometric functions of three and four
variables (see Appendix C, Two massive propagators). The solution $J_{32}$
is in accordance with the result calculated in \cite{davydy}. Note again
that there are thirty-one other solutions in different kinematical regions.

\subsection{One-loop triangle, three equal masses case}

$ $

Finally, we consider the Feynman integral with three equal masses ($
m_{1}=m_{2}=m_{3}=m$) and obtain forty solutions, which are grouped
into thirty four independent solutions (see Table below), also expressed
through the hypergeometric functions of three, four and five variables (see
Appendix D, Three massive denominators).

\vspace{.25cm}

\begin{center}
\begin{tabular}{|c|c||c|c||c|c|}
\hline
Set & Solutions & Set & Solutions & Set & Solutions \\ \hline\hline
1 & $J_1$ & 13 & $J_{16}$ & 25 & $J_{28}+J_{29}$ \\ \hline
2 & $J_2$ & 14 & $J_{17}$ & 26 & $J_{30}+J_{31}$ \\ \hline
3 & $J_3$ & 15 & $J_{18}$ & 27 & $J_{32}+J_{33}$ \\ \hline
4 & $J_4$ & 16 & $J_{19}$ & 28 & $J_{34}$ \\ \hline
5 & $J_5$ & 17 & $J_{20}$ & 29 & $J_{35}$ \\ \hline
6 & $J_6$ & 18 & $J_{21}$ & 30 & $J_{36}$ \\ \hline
7 & $J_7$ & 19 & $J_{22}$ & 31 & $J_{37}$ \\ \hline
8 & $J_8$ & 20 & $J_{23}$ & 32 & $J_{38}$ \\ \hline
9 & $J_9$ & 21 & $J_{24}$ & 33 & $J_{39}$ \\ \hline
10 & $J_{10}+J_{11}$ & 22 & $J_{25}$ & 34 & $J_{40}$ \\ \hline
11 & $J_{12}+J_{13}$ & 23 & $J_{26}$ &  & $ $ \\ \hline
12 & $J_{14}+J_{15}$ & 24 & $J_{27}$ &  & $ $ \\ \hline
\end{tabular}
\end{center}

\vspace{.25cm} The last solution, namely, $J_{40}$ is in accordance with the
result calculated in \cite{davydy}.

\section{Conclusion}

$ $

In this work we have used the NDIM approach to evaluate the one-loop
triangle diagram with massless and massive (with one, two and three equal
masses) internal particles. These kind of diagrams are relevant to, e.g.,
vertex corrections in QED and QCD, Z-scattering in electroweak interactions
and so on. For the massless case, we have three distinct kinematical regions
obtained simultaneously; for the one mass case we have sixteen distinct
kinematical regions; thirty-two for the two equal masses case and thirty-six
for the three equal masses case. All solutions are represented by
hypergeometric-type functions of several variables and some of these were
compared to existing results in the literature and in all such cases our
results do match the known ones. Therefore we deem NDIM as a powerful
technique in the computation of highly complex Feynman integrals in various
kinematical regions of interest.

\vspace{1cm}

{\bf Acknowledgments:} ATS acknowledges partial support from CNPq, Bras\'{\i}lia, Brazil; ESS wishes to thank CAPES, Bras\'{\i}lia, Brazil, and AGMS
wishes to thank FAPESP, S\~ao Paulo, Brazil and CNPq for financial support
\appendix

\section{Massless Case}

$ $

The massless solutions $J_{n}=J_{n}(a,b,c,D,p,q),$ where $n=1,2,...,12,$\
are given by Appel hypergeometric function
\[
F_{4}\left[ x_{1},x_{2};x_{3},x_{4}|z_{1};z_{2}\right] ={\sum_{i,j=0}^{
\infty }}\frac{(x_{1})_{i+j}(x_{2})_{i+j}}{(x_{3})_{i}(x_{4})_{j}}\frac{
z_{1}^{i}}{i!}\frac{z_{2}^{j}}{j!}, 
\]
and the general expression for the solutions are given by
$J_{n}=A_{n}F_{4},$
 where the coefficients $A_{n}$ are shown in
Table-1 and the parameters and
 variables of the functions $F_{4}$ in
the Table-2 below. When appropriate,
 we use double lines between two
subsequent sets of solutions in the Tables
 to separate different
kinematical regions. Also, wherever convenient, we set
$\sigma=a+b+c+D/2$. 
\[
{\ 
\begin{tabular}{|c|c|}
\hline
$n$ & $A_{n}$ \\ \hline\hline
$1$ & \multicolumn{1}{|l|}{$\pi
^{D/2}(r^{2})^{\sigma}(z_1)^{\sigma-b}(z_2)^{\sigma-c}\frac{
(-b)_{-a-D/2}(-c)_{-a-D/2}}{(a+D/2)_{-2a-D/2}}$} \\ \hline
$2$ & \multicolumn{1}{|l|}{$\pi ^{D/2}(r^2)^{\sigma}(z_2)^{\sigma-c}\frac{
(-a)_{-c+\sigma}(-b)_{\sigma}}{(c-\sigma)_{-c+2\sigma+D/2}}$} \\ \hline
$3$ & \multicolumn{1}{|l|}{$\pi ^{D/2}(r^2)^{\sigma}(z_1)^{\sigma-b}\frac{
(-a)_{-b+\sigma}(-c)_{\sigma}}{(b-\sigma)_{-b+2\sigma+D/2}}$} \\ \hline
$4$ & \multicolumn{1}{|l|}{$\pi ^{D/2}(r^{2})^{\sigma}\frac{
(-b)_{\sigma}(-c)_{\sigma}}{(-\sigma)_{2\sigma+D/2}}$} \\ \hline\hline
$5$ & \multicolumn{1}{|l|}{$\pi
^{D/2}(q^2)^{\sigma}(z_1)^{\sigma-a}(z_2)^{\sigma-c}\frac{
(-a)_{-b-D/2}(-c)_{-b-D/2}}{(b+D/2)_{-2b-D/2}}$} \\ \hline
$6$ & \multicolumn{1}{|l|}{$\pi ^{D/2}(q^{2})^{\sigma}(z_2)^{\sigma-c}\frac{
(-a)_{\sigma}(-b)_{-c+\sigma}}{(c-\sigma)_{-c+2\sigma+D/2}}$} \\ \hline
$7$ & \multicolumn{1}{|l|}{$\pi ^{D/2}(q^{2})^{\sigma}(z_1)^{\sigma-a}\frac{
(-b)_{-a+\sigma}(-c)_{\sigma}}{(a-\sigma)_{-a+2\sigma+D/2}}$} \\ \hline
$8$ & \multicolumn{1}{|l|}{$\pi ^{D/2}(q^{2})^{\sigma}\frac{
(-a)_{\sigma}(-c)_{\sigma}}{(-\sigma)_{2\sigma+D/2}}$} \\ \hline\hline
$9$ & \multicolumn{1}{|l|}{$\pi ^{D/2}(p^{2})^{\sigma}(z_1)^{\sigma-a}(z_2)^{\sigma-b}\frac{
 (-a)_{-c-D/2}(-b)_{-c-D/2}}{(c+D/2)_{-2c-D/2}}$} \\ \hline
$10$ & \multicolumn{1}{|l|}{$\pi ^{D/2}(p^{2})^{\sigma}(z_2)^{\sigma-b}\frac{
(-a)_{\sigma}(-c)_{-b+\sigma}}{(b-\sigma)_{-b+2\sigma+D/2}}$} \\ \hline
$11$ & \multicolumn{1}{|l|}{$\pi ^{D/2}(p^{2})^{\sigma}(z_1)^{\sigma-a}\frac{
(-b)_{\sigma}(-c)_{-a+\sigma}}{(a-\sigma)_{-a+2\sigma+D/2}}$} \\ \hline
$12$ & \multicolumn{1}{|l|}{$\pi ^{D/2}(p^{2})^{\sigma}\frac{
(-a)_{\sigma}(-b)_{\sigma}}{(-\sigma)_{2\sigma+D/2}}$} \\ \hline
\multicolumn{2}{c}{Table-1}
\end{tabular}
\ } 
\]
\[
{\ 
\begin{tabular}{|c|c|c|}
\hline
$n$ & \multicolumn{1}{|c|}{$x_{1},\;x_{2};\;x_{3},\;x_{4}$} & $z_{1},z_{2}$ \\ 
\hline\hline
$1$ & \multicolumn{1}{|l|}{$a+D/2,\,\sigma+D/2;\,1-b+\sigma,\,1-c+\sigma$} & $\frac{
q^{2}}{r^{2}},\frac{p^{2}}{r^{2}}$ \\ \hline
$2$ & \multicolumn{1}{|l|}{$-c,\,b+D/2;\,1+b+\sigma,\,1-c+\sigma$} & $\frac{q^{2}}{r^{2}}
,\frac{p^{2}}{r^{2}}$ \\ \hline
$3$ & \multicolumn{1}{|l|}{$-b,\,c+D/2;\,1-b+\sigma,\,1+c-\sigma$} & $\frac{q^{2}}{
r^{2}},\frac{p^{2}}{r^{2}}$ \\ \hline
$4$ & \multicolumn{1}{|l|}{$-a,\,-\sigma;\,1+b-\sigma,\,1+c-\sigma$} & $\frac{
q^{2}}{r^{2}},\frac{p^{2}}{r^{2}}$ \\ \hline\hline
$5$ & \multicolumn{1}{|l|}{$b+D/2,\,\sigma+D/2;\,1-a+\sigma,\,1-c+\sigma$} & $\frac{
r^{2}}{q^{2}},\frac{p^{2}}{q^{2}}$ \\ \hline
$6$ & \multicolumn{1}{|l|}{$-c,\,a+D/2;\,1+a-\sigma,\,1-c+\sigma$} & $\frac{r^{2}}{
q^{2}},\frac{p^{2}}{q^{2}}$ \\ \hline
$7$ & \multicolumn{1}{|l|}{$-a,\,c+D/2;\,1-a+\sigma,\,1+c-\sigma$} & $\frac{r^{2}}{
q^{2}},\frac{p^{2}}{q^{2}}$ \\ \hline
$8$ & \multicolumn{1}{|l|}{$-b,\,-\sigma;\,,1+a-\sigma,\,1+c-\sigma$} & $\frac{
r^{2}}{q^{2}},\frac{p^{2}}{q^{2}}$ \\ \hline\hline
$9$ & \multicolumn{1}{|l|}{$c+D/2,\,\sigma+D/2;\,1-a+\sigma,\,1-b+\sigma$} & $\frac{
r^{2}}{p^{2}},\frac{q^{2}}{p^{2}}$ \\ \hline
$10$ & \multicolumn{1}{|l|}{$-b,\,a+D/2;\,1+a-\sigma,\,1-b+\sigma$} & $\frac{r^{2}}{
p^{2}},\frac{q^{2}}{p^{2}}$ \\ \hline
$11$ & \multicolumn{1}{|l|}{$-a,\,b+D/2;\,1-a+\sigma,\,1+b-\sigma$} & $\frac{r^{2}}{
p^{2}},\frac{q^{2}}{p^{2}}$ \\ \hline
$12$ & \multicolumn{1}{|l|}{$-c,\,-\sigma;\,1+a-\sigma,\,1+b-\sigma$} & $\frac{
r^{2}}{p^{2}},\frac{q^{2}}{p^{2}}$ \\ \hline
\multicolumn{2}{c}{Table-2}  
\end{tabular}
\ } 
\]

\section{One massive propagator}

$ $

The one massive denominator solutions $J_{n}=J_{n}(a,b,c,D,p,q,m,0,0),$
where $n=1,2,...,23,$ are given by hypergeometric functions listed below 
\begin{eqnarray*}
T_{1}\left[ 
\begin{array}{l|}
x_{1},x_{2},x_{3} \\ 
x_{4},x_{5}
\end{array}
\;\,z_{1};z_{2};z_{3}\right] &=&{\sum_{j_{1},j_{2},j_{3}=0}^{\infty }}\frac{(x_{1})_{j_{1}+j_{2}+j_{3}}(x_{2})_{j_{1}+j_{3}}(x_{3})_{j_{2}+j_{3}}}{(x_{4})_{j_{1}+j_{2}+j_{3}}(x_{5})_{j_{3}}}\frac{z_{1}^{j_{1}}}{j_{1}!}\frac{z_{2}^{j_{2}}}{j_{2}!}\frac{z_{3}^{j_{3}}}{j_{3}!}, \\
T_{2}\left[ 
\begin{array}{l|}
x_{1},x_{2},x_{3} \\ 
x_{4},x_{5}
\end{array}
\;\,z_{1};z_{2};z_{3}\right] &=&{\sum_{j_{1},j_{2},j_{3}=0}^{\infty }}\frac{
(x_{1})_{j_{1}+j_{2}-j_{3}}(x_{2})_{j_{1}+j_{2}-j_{3}}(x_{3})_{j_{3}}}{
(x_{4})_{j_{1}-j_{3}}(x_{5})_{j_{2}-j_{3}}}\frac{z_{1}^{j_{1}}}{j_{1}!}\frac{
z_{2}^{j_{2}}}{j_{2}!}\frac{z_{3}^{j_{3}}}{j_{3}!}, \\
T_{3}\left[ 
\begin{array}{l|}
x_{1},x_{2},x_{3} \\ 
x_{4},x_{5}
\end{array}
\;\,z_{1};z_{2};z_{3}\right] &=&{\sum_{j_{1},j_{2},j_{3}=0}^{\infty }}\frac{
(x_{1})_{j_{1}+j_{2}-j_{3}}(x_{2})_{j_{1}+j_{2}}(x_{3})_{j_{3}}}{
(x_{4})_{j_{2}-j_{3}}(x_{5})_{j_{1}}}\frac{z_{1}^{j_{1}}}{j_{1}!}\frac{
z_{2}^{j_{2}}}{j_{2}!}\frac{z_{3}^{j_{3}}}{j_{3}!}, \\
T_{4}\left[ 
\begin{array}{l|}
x_{1},x_{2},x_{3} \\ 
x_{4},x_{5}
\end{array}
\;\,z_{1};z_{2};z_{3}\right] &=&{\sum_{j_{1},j_{2},j_{3}=0}^{\infty }}\frac{
(x_{1})_{j_{1}+j_{2}}(x_{2})_{j_{1}+j_{2}}(x_{3})_{j_{1}-j_{3}}}{
(x_{4})_{j_{1}-j_{3}}(x_{5})_{j_{1}+j_{2}-j_{3}}}\frac{z_{1}^{j_{1}}}{j_{1}!}
\frac{z_{2}^{j_{2}}}{j_{2}!}\frac{z_{3}^{j_{3}}}{j_{3}!}, \\
T_{5}\left[ 
\begin{array}{l|}
x_{1},x_{2},x_{3} \\ 
x_{4},x_{5}
\end{array}
\;\,z_{1};z_{2};z_{3}\right] &=&{\sum_{j_{1},j_{2},j_{3}=0}^{\infty }}\frac{
(x_{1})_{j_{1}+j_{2}}(x_{2})_{j_{1}+j_{3}}(x_{3})_{j_{2}-j_{3}}}{
(x_{4})_{j_{2}-j_{3}}(x_{5})_{j_{1}}}\frac{z_{1}^{j_{1}}}{j_{1}!}\frac{
z_{2}^{j_{2}}}{j_{2}!}\frac{z_{3}^{j_{3}}}{j_{3}!}, \\
T_{6}\left[ 
\begin{array}{l|}
x_{1},x_{2},x_{3} \\ 
x_{4},x_{5}
\end{array}
\;\,z_{1};z_{2};z_{3}\right] &=&{\sum_{j_{1},j_{2},j_{3}=0}^{\infty }}\frac{
(x_{1})_{j_{1}+j_{2}}(x_{2})_{j_{1}+j_{2}}(x_{3})_{j_{3}}}{
(x_{4})_{j_{2}-j_{3}}(x_{5})_{j_{1}+j_{3}}}\frac{z_{1}^{j_{1}}}{j_{1}!}\frac{
z_{2}^{j_{2}}}{j_{2}!}\frac{z_{3}^{j_{3}}}{j_{3}!}, \\
T_{7}\left[ 
\begin{array}{l|}
x_{1},x_{2},x_{3} \\ 
x_{4},x_{5}
\end{array}
\;\,z_{1};z_{2};z_{3}\right] &=&{\sum_{j_{1},j_{2},j_{3}=0}^{\infty }}\frac{
(x_{1})_{j_{1}+j_{2}+j_{3}}(x_{2})_{j_{1}+j_{2}}(x_{3})_{j_{3}}}{
(x_{4})_{j_{1}+j_{3}}(x_{5})_{j_{2}}}\frac{z_{1}^{j_{1}}}{j_{1}!}\frac{
z_{2}^{j_{2}}}{j_{2}!}\frac{z_{3}^{j_{3}}}{j_{3}!}, \\
T_{8}\left[ 
\begin{array}{l|}
x_{1},x_{2},x_{3} \\ 
x_{4},x_{5}
\end{array}
\;\,z_{1};z_{2};z_{3}\right] &=&{\sum_{j_{1},j_{2},j_{3}=0}^{\infty }}\frac{
(x_{1})_{j_{1}+j_{2}+j_{3}}(x_{2})_{j_{1}+j_{2}+j_{3}}(x_{3})_{j_{3}}}{
(x_{4})_{j_{2}+j_{3}}(x_{5})_{j_{1}+j_{3}}}\frac{z_{1}^{j_{1}}}{j_{1}!}\frac{
z_{2}^{j_{2}}}{j_{2}!}\frac{z_{3}^{j_{3}}}{j_{3}!},
\end{eqnarray*}
and the expression of each one solution is given by $J_{n}=B_{n}T_{l},$
where the relation between $n$ and $l$ is given by 
\[
\begin{tabular}{|c|c|c|c|c|c|c|c|c|}
\hline
$n$ & $1,2,3$ & $4$ & $5,\cdots,8$ & $9,\cdots,12$ & $13,\cdots,16$ & $17,\;18$
& $19,\cdots,22$ & $23$ \\ \hline
$l$ & $1$ & $2$ & $3$ & $4$ & $5$ & $6$ & $7$ & $8$ \\ \hline
\end{tabular}
\]
where the coefficients $B_{n}$ are shown in Table-3, the parameters\ and
variables of the functions $T_{l}$ in the Table-4 below (consider $\sigma
=a+b+c+D/2$). 
\[
{\ 
\begin{tabular}{|c|c|}
\hline
$n$ & $B_{n}$ \\ \hline\hline
$1$ & \multicolumn{1}{|l|}{$\pi ^{D/2}(-m^{2})^{\sigma}(z_3)^{\sigma-a}\frac{
(-b)_{-a+\sigma}(-c)_{-a+\sigma}}{(a-\sigma)_{-2a+2\sigma+D/2}}$} \\ \hline
$2$ & \multicolumn{1}{|l|}{$\pi ^{D/2}(-m^{2})^{\sigma}\frac {(-a)_{\sigma}}{(-\sigma)_{\sigma+D/2}}$} \\ \hline\hline
$3$ & \multicolumn{1}{|l|}{$\pi^ {D/2}(-m^{2})^{\sigma}(-z_1)^{-b}(-z_2)^{-c}\frac {(-a)_{\sigma}}{(-a-D/2)_{\sigma+D/2}}$} \\ \hline\hline
$4$ & \multicolumn{1}{|l|}{$\pi^ {D/2}(r^{2})^{\sigma}(z_1)^{\sigma-b}(z_2)^{\sigma-c}\frac {(-b)_{-a-D/2}(-c)_{-a-D/2}}{(a+D/2)_{-2a-D/2}} 
$} \\ \hline\hline
$5$ & \multicolumn{1}{|l|}{$\pi^ {D/2}(q^{2})^{\sigma}(z_1)^{\sigma-a}(z_2)^{\sigma-c}\frac {(-a)_{-b-D/2}(-c)_{-b-D/2}}{(b+D/2)_{-2b-D/2}} 
$} \\ \hline
$6$ & \multicolumn{1}{|l|}{$\pi ^{D/2}(q^{2})^{\sigma}(z_2)^{\sigma-c}\frac{
(-a)_{\sigma}(-b)_{-c+\sigma}}{(c-\sigma)_{-c+2\sigma+D/2}}$} \\ \hline\hline
$7$ & \multicolumn{1}{|l|}{$\pi^ {D/2}(p^{2})^{\sigma}(z_1)^{\sigma-a}(z_2)^{\sigma-b}\frac {(-a)_{-c-D/2}(-b)_{-c-D/2}}{(c+D/2)_{-2c-D/2}}$} \\ \hline
$8$ & \multicolumn{1}{|l|}{$\pi ^{D/2}(p^{2})^{\sigma}(z_2)^{\sigma-b}\frac{
(-a)_{\sigma}(-c)_{-b+\sigma}}{(b-\sigma)_{-b+2\sigma+D/2}}$} \\ \hline\hline
$9$ & \multicolumn{1}{|l|}{$\pi^ {D/2}(p^{2})^{\sigma}(z_2)^{-c}(z_3)^{\sigma-b}\frac{(-a)_{-c+\sigma}(-b)_{c}}{(b-\sigma)_{\sigma+D/2}}$} \\ 
\hline\hline
$10$ & \multicolumn{1}{|l|}{$\pi^ {D/2}(q^{2})^{\sigma}(z_2)^{-b}(z_3)^{\sigma-c}\frac{(-a)_{-b+\sigma}(-c)_{b}}{(c-\sigma)_{\sigma+D/2}}$} \\ 
\hline\hline
$11$ & \multicolumn{1}{|l|}{$\pi ^{D/2}(r^{2})^{\sigma}(z_1)^{\sigma-c}\frac{(-a)_{-c+\sigma}(-b)_{c}}{(c-\sigma)_{\sigma+D/2}}$} \\ \hline\hline
$12$ & \multicolumn{1}{|l|}{$\pi ^{D/2}(r^{2})^{\sigma}(z_1)^{\sigma-b}\frac{(-a)_{-b+\sigma}(-c)_{b}}{(b-\sigma)_{\sigma+D/2}}$} \\ \hline\hline
$13$ & \multicolumn{1}{|l|}{$\pi^ {D/2}(p^{2})^{\sigma}(z_1)^{\sigma-a}(-z_2)^{\sigma-b}\frac{(-a)_{-c-D/2}(-b)_{-c-D/2}}{(c+D/2)_{-2c-D/2}} $} \\ \hline
$14$ & \multicolumn{1}{|l|}{$\pi^ {D/2}(p^{2})^{\sigma}(-z_2)^{\sigma-b}\frac{(-a)_{\sigma}}{(b-\sigma)_{\sigma+D/2}}$} \\ \hline\hline
$15$ & \multicolumn{1}{|l|}{$\pi^ {D/2}(q^{2})^{\sigma}(z_1)^{\sigma-a}(-z_2)^{\sigma-c}\frac{(-a)_{-b-D/2}(-c)_{-b-D/2}}{(b+D/2)_{-2b-D/2}} $} \\ \hline
$16$ & \multicolumn{1}{|l|}{$\pi^ {D/2}(q^{2})^{\sigma}(-z_2)^{\sigma-c}\frac{(-a)_{\sigma}}{(c-\sigma)_{\sigma+D/2}}$} \\ \hline\hline
$17$ & \multicolumn{1}{|l|}{$\pi^ {D/2}(r^{2})^{\sigma}(z_2)^{\sigma-c}\frac{(-a)_{-c+\sigma}(-b)_{\sigma}}{(c-\sigma)_{-c+2\sigma+D/2}}$} \\ \hline\hline
$18$ & \multicolumn{1}{|l|}{$\pi ^{D/2}(r^{2})^{\sigma}(z_2)^{\sigma-b}\frac{(-a)_{-b+\sigma}(-c)_{\sigma}}{(b-\sigma)_{-b+2\sigma+D/2}}$} \\ \hline\hline
$19$ & \multicolumn{1}{|l|}{$\pi ^{D/2}(p^{2})^{\sigma}(z_2)^{\sigma-a}\frac{(-b)_{\sigma}(-c)_{-a+\sigma}}{(a-\sigma)_{-a+2\sigma+D/2}}$} \\ \hline
$20$ & \multicolumn{1}{|l|}{$\pi ^{D/2}(p^{2})^{\sigma}\frac{
(-a)_{\sigma}(-b)_{\sigma}}{(-\sigma)_{2\sigma+D/2}}$} \\ \hline\hline
$21$ & \multicolumn{1}{|l|}{$\pi ^{D/2}(q^{2})^{\sigma}(z_2)^{\sigma-a}\frac{
(-b)_{-a+\sigma}(-c)_{\sigma}}{(a-\sigma)_{-a+2\sigma+D/2}}$} \\ \hline
$22$ & \multicolumn{1}{|l|}{$\pi ^{D/2}(q^{2})^{\sigma}\frac{
(-a)_{\sigma}(-c)_{\sigma}}{(-\sigma)_{2\sigma+D/2}}$} \\ \hline\hline
$23$ & \multicolumn{1}{|l|}{$\pi ^{D/2}(r^{2})^{\sigma}\frac{
(-b)_{\sigma}(-c)_{\sigma}}{(-\sigma)_{2\sigma+D/2}}$} \\ \hline
\multicolumn{2}{c}{Table-3}
\end{tabular}
\ } 
\]

\[
{\ 
\begin{tabular}{|c|l|l|}
\hline
$n$ & \multicolumn{1}{|c|}{$x_{1},\;x_{2},\;x_{3};\:x_{4},\;x_{5}$} & 
\multicolumn{1}{|c|}{$z_{1},\;z_{2},\;z_{3}$} \\ \hline\hline
$1$ & $-a,\,c+D/2,\,b+D/2;\;\sigma-a+D/2,\,1-a+\sigma$ & $\frac{p^{2}}{m^{2}};\,\frac{q^{2}}{
m^{2}};\,-\frac{r^{2}}{m^{2}}$ \\ \hline
$2$ & $-\sigma,\,-b,\,-c;\;D/2,\,1+a-\sigma$ & $\frac{p^{2}}{m^{2}};\,\frac{q^{2}}{
m^{2}};\,-\frac{r^{2}}{m^{2}}$ \\ \hline\hline
$3$ & $1+a-\sigma,\,-b,\,-c;\,1+a+D/2,\,1+a-\sigma$ & $\frac{m^{2}}{p^{2}};\;\frac{
m^{2}}{q^{2}};\;-\frac{m^{2}r^{2}}{p^{2}q^{2}}$ \\ \hline\hline
$4$ & $a+D/2,\,\sigma+D/2,\,1-\sigma-D/2;\,1-b+\sigma,\,1-c+\sigma$ & $\frac{
q^{2}}{r^{2}};\;\frac{p^{2}}{r^{2}};\;-\frac{m^{2}r^{2}}{p^{2}q^{2}}$ \\ 
\hline\hline
$5$ & $\sigma+D/2,\,b+D/2,\,1-\sigma-D/2;1-c+\sigma,\,1-a+\sigma$ & $\frac{r^{2}}{q^{2}};\;\frac{p^{2}}{q^{2}};\;\frac{m^{2}}{p^{2}}$ \\ \hline
$6$ & $a+D/2,\,-c,\,1-\sigma-D/2;\,1-c+\sigma,\,1+a-\sigma$ & $\frac{r^{2}}{q^{2}};\;\frac{p^{2}}{q^{2}};\;\frac {m^2}{p^2}$ \\ \hline\hline
$7$ & $\sigma+D/2,\,c+D/2,\,1-\sigma-D/2;\,1-b+\sigma,1-a+\sigma$ & $\frac{
r^{2}}{p^{2}};\;\frac{q^{2}}{p^{2}};\;\frac{m^{2}}{q^{2}}$ \\ \hline
$8$ & $a+D/2,\,-b,\,1-\sigma -D/2;\,1-b+\sigma,1+a-\sigma$ & $\frac{r^{2}}{p^{2}};\;\frac{q^{2}}{p^{2}};\;\frac{m^{2}}{q^{2}}$ \\ \hline\hline
$9$ & $-c,\,b+D/2,\,b-\sigma;\,b+D/2,\,1+b-c$ & $-\frac{p^{2}q^{2}}{m^{2}r^{2}};\;
\frac{p^{2}}{r^{2}};\;-\frac{m^{2}}{p^{2}}$ \\ \hline\hline
$10$ & $-b,\,c+D/2,\,c-\sigma;\,c+D/2,\,1-b+c$ & $-\frac{p^{2}q^{2}}{m^{2}r^{2}};\;
\frac{q^{2}}{r^{2}};\;-\frac{m^{2}}{q^{2}}$ \\ \hline\hline
$11$ & $-c,\,b+D/2,\,1-c-D/2;\,1-c+\sigma,\,1+b-c$ & $-\frac{m^{2}}{r^{2}};\;\frac{
q^{2}}{r^{2}};\;-\frac{p^{2}}{m^{2}}$ \\ \hline\hline
$12$ & $-b,\,c+D/2,\,1-b-D/2;\,1-b+\sigma,\,1-b+c$ & $-\frac{m^{2}}{r^{2}};\;\frac{
p^{2}}{r^{2}};\;-\frac{q^{2}}{m^{2}}$ \\ \hline\hline
$13$ & $c+D/2,\,b+D/2,\,1-b-D/2;\,1-b+\sigma,\,1-a+\sigma $ & $\frac{r^{2}}{p^{2}};\;\frac{m^{2}}{p^{2}};\;\frac{q^{2}}{m^{2}}$ \\ \hline
$14$ & $-b,\,-c,\,1-b-D/2;\,1-b+\sigma,\,1+a-\sigma $ & $\frac{r^{2}}{p^{2}};\;\frac{m^{2}}{p^{2}};\;\frac{q^{2}}{m^{2}}$ \\ \hline\hline
$15$ & $b+D/2,\,c+D/2,\,1-c-D/2;\,1-c+\sigma,\,1-a+\sigma $ & $\frac{r^{2}}{q^{2}};\;\frac{m^{2}}{q^{2}};\;\frac{p^{2}}{m^{2}}$ \\ \hline
$16$ & $-c,\,-b,\,1-c-D/2;\,1-c+\sigma,\,1+a-\sigma $ & $\frac{r^{2}}{q^{2}};\;\frac{m^{2}}{q^{2}};\;\frac{p^{2}}{m^{2}}$ \\ \hline\hline
$17$ & $-c,\,b+D/2,\,1-\sigma-D/2;\,1-c+\sigma,\,1+b-\sigma $ & $\frac{q^{2}}{r^{2}};\;\frac{p^{2}}{r^{2}};\;-\frac{m^{2}}{p^{2}}$ \\ \hline\hline$18$ & $-b,\,c+D/2,\,1-\sigma-D/2;\,1-b+\sigma,\,1+c-\sigma $ & $\frac{p^{2}}{r^{2}};\;\frac{q^{2}}{r^{2}};\;-\frac{m^{2}}{q^{2}}$ \\ \hline\hline
$19$ & $-a,\,b+D/2,\,1-\sigma-D/2;\,1+b-\sigma,\,1-a+\sigma $ & $\frac{q^{2}}{p^{2}};\;\frac{r^{2}}{p^{2}};\;\frac{m^{2}}{p^{2}}$ \\ \hline
$20$ & $-\sigma,\,-c,\,1-\sigma-D/2;\,1+b-\sigma,\,1+a-\sigma $ & $\frac{q^{2}}{p^{2}};\;\frac{r^{2}}{p^{2}};\;\frac{m^{2}}{p^{2}}$ \\ \hline\hline
$21$ & $-a,\,c+D/2,\,1-\sigma-D/2;\,1+c-\sigma,\,1-a+\sigma $ & $\frac{p^{2}}{q^{2}};\;\frac{r^{2}}{q^{2}};\;\frac{m^{2}}{q^{2}}$ \\ \hline
$22$ & $-\sigma,\,-b,\,1-\sigma-D/2;\,1+c-\sigma,\,1+a-\sigma $ & $\frac{p^{2}}{q^{2}};\;\frac{r^{2}}{q^{2}};\;\frac{m^{2}}{q^{2}}$ \\ \hline\hline
$23$ & $-a,\,-\sigma,\,1-\sigma-D/2;\,1+b-\sigma,\,1+c-\sigma $ & $\frac{p^{2}}{r^{2}};\;\frac{q^{2}}{r^{2}};\;-\frac{m^{2}}{r^{2}}$ \\ \hline
\multicolumn{2}{c}{Table-4} 
\end{tabular}
}. 
\]

\section{Two massive propagators}

$ $

The two massive denominators solutions $J_{n}=J_{n}(a,b,c,D,p,q,0,m,m),$
where $n=1,2,...,32,$ are given by hypergeometric functions listed below 
\begin{eqnarray*}
R_{1}\left[ 
\begin{array}{l|}
x_{1},x_{2},x_{3} \\ 
x_{4},x_{5}
\end{array}
\;\,z_{1};z_{2};z_{3};z_{4}\right]  &=&{\sum_{j_{1},..,j_{4}=0}^{\infty }}\frac{
(x_{1})_{j_{1}+j_{2}+j_{3}}(x_{2})_{j_{1}+j_{2}-j_{4}}(x_{3})_{j_{2}+j_{3}}}{
(x_{4})_{j_{2}-j_{4}}(x_{5})_{j_{1}+j_{2}+j_{3}-j_{4}}}\frac{z_{1}^{j_{1}}}{
j_{1}!}\frac{z_{2}^{j_{2}}}{j_{2}!}\frac{z_{3}^{j_{3}}}{j_{3}!}\frac{
z_{4}^{j_{4}}}{j_{4}!}, \\
R_{2}\left[ 
\begin{array}{l|}
x_{1},x_{2},x_{3} \\ 
x_{4},x_{5}
\end{array}
\;\,z_{1};z_{2};z_{3};z_{4}\right]  &=&{\sum_{j_{1},..,j_{4}=0}^{\infty }}\frac{
(x_{1})_{j_{1}+j_{2}-j_{3}-j_{4}}(x_{2})_{j_{1}+j_{2}}(x_{3})_{j_{3}+j_{4}}}{
(x_{4})_{j_{1}-j_{3}}(x_{5})_{j_{2}-j_{4}}}\frac{z_{1}^{j_{1}}}{j_{1}!}\frac{
z_{2}^{j_{2}}}{j_{2}!}\frac{z_{3}^{j_{3}}}{j_{3}!}\frac{z_{4}^{j_{4}}}{j_{4}!
}, \\
R_{3}\left[ 
\begin{array}{l|}
x_{1},x_{2},x_{3} \\ 
x_{4},x_{5}
\end{array}
\;\,z_{1};z_{2};z_{3};z_{4}\right]  &=&{\sum_{j_{1},..,j_{4}=0}^{\infty }}\frac{
(x_{1})_{j_{1}+j_{2}-j_{3}-j_{4}}(x_{2})_{j_{1}+j_{2}-j_{3}}(x_{3})_{j_{3}+j_{4}}
}{(x_{4})_{j_{2}-j_{3}}(x_{5})_{j_{1}-j_{3}-j_{4}}}\frac{z_{1}^{j_{1}}}{
j_{1}!}\frac{z_{2}^{j_{2}}}{j_{2}!}\frac{z_{3}^{j_{3}}}{j_{3}!}\frac{
z_{4}^{j_{4}}}{j_{4}!}, \\
R_{4}\left[ 
\begin{array}{l|}
x_{1},x_{2},x_{3} \\ 
x_{4},x_{5}
\end{array}
\;\,z_{1};z_{2};z_{3};z_{4}\right]  &=&{\sum_{j_{1},..,j_{4}=0}^{\infty }}\frac{
(x_{1})_{j_{1}+j_{2}-j_{3}}(x_{2})_{j_{1}+j_{4}}(x_{3})_{j_{2}-j_{4}}}{
(x_{4})_{j_{2}-j_{3}-j_{4}}(x_{5})_{j_{1}-j_{3}}}\frac{z_{1}^{j_{1}}}{j_{1}!}
\frac{z_{2}^{j_{2}}}{j_{2}!}\frac{z_{3}^{j_{3}}}{j_{3}!}\frac{z_{4}^{j_{4}}}{
j_{4}!}, \\
R_{5}\left[ 
\begin{array}{l|}
x_{1},x_{2},x_{3} \\ 
x_{4},x_{5}
\end{array}
\;\,z_{1};z_{2};z_{3};z_{4}\right]  &=&{\sum_{j_{1},..,j_{4}=0}^{\infty }}\frac{
(x_{1})_{j_{1}+j_{2}+j_{3}}(x_{2})_{j_{1}+j_{3}}(x_{3})_{j_{2}+j_{4}}}{
(x_{4})_{j_{1}+j_{2}+j_{4}}(x_{5})_{j_{3}-j_{4}}}\frac{z_{1}^{j_{1}}}{j_{1}!}
\frac{z_{2}^{j_{2}}}{j_{2}!}\frac{z_{3}^{j_{3}}}{j_{3}!}\frac{z_{4}^{j_{4}}}{
j_{4}!}, \\
R_{6}\left[ 
\begin{array}{l|}
x_{1},x_{2},x_{3} \\ 
x_{4},x_{5}
\end{array}
\;\,z_{1};z_{2};z_{3};z_{4}\right]  &=&{\sum_{j_{1},..,j_{4}=0}^{\infty }}\frac{
(x_{1})_{j_{1}+j_{2}+j_{3}+j_{4}}(x_{2})_{j_{1}+j_{2}+j_{3}}(x_{3})_{j_{1}+j_{4}}
}{(x_{4})_{j_{1}+j_{3}+j_{4}}(x_{5})_{j_{1}+j_{2}}}\frac{z_{1}^{j_{1}}}{
j_{1}!}\frac{z_{2}^{j_{2}}}{j_{2}!}\frac{z_{3}^{j_{3}}}{j_{3}!}\frac{
z_{4}^{j_{4}}}{j_{4}!}, \\
R_{7}\left[ 
\begin{array}{l|}
x_{1},x_{2},x_{3} \\ 
x_{4},x_{5}
\end{array}
\;\,z_{1};z_{2};z_{3};z_{4}\right]  &=&{\sum_{j_{1},..,j_{4}=0}^{\infty }}\frac{
(x_{1})_{j_{1}+j_{2}+j_{3}}(x_{2})_{j_{1}+j_{2}-j_{4}}(x_{3})_{j_{3}+j_{4}}}{
(x_{4})_{j_{2}-j_{4}}(x_{5})_{j_{1}+j_{3}}}\frac{z_{1}^{j_{1}}}{j_{1}!}\frac{
z_{2}^{j_{2}}}{j_{2}!}\frac{z_{3}^{j_{3}}}{j_{3}!}\frac{z_{4}^{j_{4}}}{j_{4}!
}, \\
R_{8}\left[ 
\begin{array}{l|}
x_{1},x_{2},x_{3} \\ 
x_{4},x_{5}
\end{array}
\;\,z_{1};z_{2};z_{3};z_{4}\right]  &=&{\sum_{j_{1},..,j_{4}=0}^{\infty }}\frac{
(x_{1})_{j_{1}+j_{2}-j_{3}}(x_{2})_{j_{1}+j_{2}}(x_{3})_{j_{3}+j_{4}}}{
(x_{4})_{j_{1}-j_{3}-j_{4}}(x_{5})_{j_{2}+j_{4}}}\frac{z_{1}^{j_{1}}}{j_{1}!}
\frac{z_{2}^{j_{2}}}{j_{2}!}\frac{z_{3}^{j_{3}}}{j_{3}!}\frac{z_{4}^{j_{4}}}{
j_{4}!}, \\
R_{9}\left[ 
\begin{array}{l|}
x_{1},x_{2},x_{3} \\ 
x_{4},x_{5}
\end{array}
\;\,z_{1};z_{2};z_{3};z_{4}\right]  &=&{\sum_{j_{1},..,j_{4}=0}^{\infty }}\frac{
(x_{1})_{j_{1}+j_{2}+j_{3}+j_{4}}(x_{2})_{j_{1}+j_{2}}(x_{3})_{j_{3}+j_{4}}}{
(x_{4})_{j_{1}+j_{3}}(x_{5})_{j_{2}+j_{4}}}\frac{z_{1}^{j_{1}}}{j_{1}!}\frac{
z_{2}^{j_{2}}}{j_{2}!}\frac{z_{3}^{j_{3}}}{j_{3}!}\frac{z_{4}^{j_{4}}}{j_{4}!
}, \\
T_{9}\left[ 
\begin{array}{c|}
x_{1},x_{2},x_{3},x_{4},x_{5} \\ 
x_{6},x_{7}
\end{array}
\;\,z_{1},z_{2};z_{3}\right]  &=&{\sum_{j_{1},j_{2},j_{3}=0}^{\infty }}\frac{
(x_{1})_{j_{1}+j_{2}}(x_{2})_{j_{2}+j_{3}}(x_{3})_{-j_{1}+j_{3}}(x_{4})_{-2j_{1}+j_{3}}(x_{5})_{j_{1}}
}{(x_{6})_{-j_{1}+j_{2}+j_{3}}(x_{7})_{-j_{1}+j_{3}}} \\
&&\times \frac{z_{1}^{j_{1}}}{j_{1}!}\frac{z_{2}^{j_{2}}}{j_{2}!}\frac{
z_{3}^{j_{3}}}{j_{3}!}, \\
T_{10}\left[ 
\begin{array}{c|}
x_{1},x_{2},x_{3},x_{4},x_{5} \\ 
x_{6},x_{7}
\end{array}
\;\,z_{1},z_{2};z_{3}\right]  &=&{\sum_{j_{1},j_{2},j_{3}=0}^{\infty }}\frac{
(x_{1})_{j_{1}+j_{2}+j_{3}}(x_{2})_{j_{1}+j_{2}}(x_{3})_{j_{1}+j_{3}}(x_{4})_{j_{2}+j_{3}}(x_{5})_{j_{3}}
}{(x_{6})_{j_{1}+j_{2}+j_{3}}(x_{7})_{j_{1}+j_{2}+2j_{3}}} \\
&&\times \frac{z_{1}^{j_{1}}}{j_{1}!}\frac{z_{2}^{j_{2}}}{j_{2}!}\frac{
z_{3}^{j_{3}}}{j_{3}!}
\end{eqnarray*}
and the expression for each one solution is given by $J_{n}=C_{n}R_{l}$, $
J_{30}=C_{30}T_{9}$, $J_{31}=C_{31}T_{9}$, $J_{32}=C_{32}T_{10}$ where the
relation between $n$ and $l$ is given by 
\[
\begin{tabular}{|c|c|c|c|c|c|c|c|c|c|}
\hline
$n$ & $1,\cdots,6$ & $7,8$ & $9,\cdots,12$ & $13,\cdots,16$ & $17,18$ & $
19,\cdots,22$ & $23,\cdots,26$ & $27,28$ & $29$ \\ \hline
$l$ & $1$ & $2$ & $3$ & $4$ & $5$ & $6$ & $7$ & $8$ & $9$ \\ \hline
\end{tabular}
\ 
\]
where the coefficients $C_{n}$ are shown in Table-5 and the parameters and
variables of the $R_{l}$\ and $T_{l}$ functions in the Table-6 below. 
\[
\begin{tabular}{|r|c|}
\hline
$n$ & $C_{n}$ \\ \hline\hline
$1$ & \multicolumn{1}{|l|}{$\pi^{D/2}(p^{2})^{\sigma}\frac{(-a)_{\sigma}(-b)_{\sigma}}{(-\sigma)_{2\sigma+D/2}}$} \\ \hline
$2$ & \multicolumn{1}{|l|}{$\pi^{D/2}(q^{2})^{\sigma}\frac{(-a)_{\sigma}(-c)_{\sigma}}{(-\sigma)_{2\sigma+D/2}}$} \\ \hline
$3$ & \multicolumn{1}{|l|}{$\pi^{D/2}(r^{2})^{\sigma}\frac{(-b)_{\sigma}(-c)_{\sigma}}{(-\sigma)_{2\sigma+D/2}}$} \\ \hline
$4$ & \multicolumn{1}{|l|}{$\pi^{D/2}(p^{2})^{\sigma}(z_2)^{-c}\frac{(-a)_{-c+\sigma}(-b)_{\sigma}}{(c-\sigma)_{-c+2\sigma+D/2}}$} \\ \hline
$5$ & \multicolumn{1}{|l|}{$\pi^{D/2}(q^{2})^{\sigma}(z_2)^{-b}\frac{(-a)_{-b+\sigma}(-c)_{\sigma}}{(b-\sigma)_{-b+2\sigma+D/2}}$} \\ \hline
$6$ & \multicolumn{1}{|l|}{$\pi^{D/2}(p^{2})^{\sigma}(z_3)^{-c}\frac{(-a)_{\sigma}(-b)_{-c+\sigma}}{(c-\sigma)_{-c+2\sigma+D/2}}$} \\ \hline
$7$ & \multicolumn{1}{|l|}{$\pi^{D/2}(q^{2})^{\sigma}(z_3)^{-b}\frac{(-a)_{\sigma}(-c)_{-b+\sigma}}{(b-\sigma)_{-b+2\sigma+D/2}}$} \\ \hline
$8$ & \multicolumn{1}{|l|}{$\pi^{D/2}(r^{2})^{\sigma}(z_1)^{-a}\frac{(-b)_{-a+\sigma}(-c)_{\sigma}}{(a-\sigma)_{-a+2\sigma+D/2}}$} \\ \hline
$9$ & \multicolumn{1}{|l|}{$\pi^{D/2}(r^{2})^{\sigma}(z_1)^{-a}\frac{(-b)_{\sigma}(-c)_{-a+\sigma}}{(a-\sigma)_{-a+2\sigma+D/2}}$} \\ \hline
$10$ & \multicolumn{1}{|l|}{$\pi^{D/2}(p^{2})^{\sigma}(z_1)^{\sigma-a}(z_2)^{\sigma-b}\frac{(-a)_{-c-D/2}(-b)_{-c-D/2}}{(c+D/2)_{-2c-D/2}}$} \\ \hline
$11$ & \multicolumn{1}{|l|}{$\pi^{D/2}(q^{2})^{\sigma}(z_1)^{\sigma-a}(z_2)^{\sigma-c}\frac{(-a)_{-b-D/2}(-c)_{-b-D/2}}{(b+D/2)_{-2b-D/2}}$} \\ \hline
$12$ & \multicolumn{1}{|l|}{$\pi^{D/2}(r^{2})^{\sigma}(z_1)^{\sigma-b}(z_2)^{\sigma-c}\frac{(-b)_{-a-D/2}(-c)_{-a-D/2}}{(a+D/2)_{-2a-D/2}}$} \\ \hline
$13$ & \multicolumn{1}{|l|}{$\pi^{D/2}(p^{2})^{\sigma}(z_1)^{\sigma-b}(-z_2)^{\sigma-a}\frac{(-a)_{-c-D/2}(-b)_{-c-D/2}}{(c+D/2)_{-2c-D/2}}$} \\ \hline
$14$ & \multicolumn{1}{|l|}{$\pi^{D/2}(q^{2})^{\sigma}(z_1)^{\sigma-c}(-z_2)^{\sigma-a}\frac{(-a)_{-b-D/2}(-c)_{-b-D/2}}{(b+D/2)_{-2b-D/2}}$} \\ \hline
$15$ & \multicolumn{1}{|l|}{$\pi^{D/2}(r^{2})^{\sigma}(z_1)^{\sigma-c}(-z_4)^{\sigma-b}\frac{(-b)_{-a-D/2}(-c)_{-a-D/2}}{(a+D/2)_{-2a-D/2}}$} \\ \hline
$16$ & \multicolumn{1}{|l|}{$\pi^{D/2}(r^{2})^{\sigma}(z_1)^{\sigma-b}(-z_4)^{\sigma-c}\frac{(-b)_{-a-D/2}(-c)_{-a-D/2}}{(a+D/2)_{-2a-D/2}}
$} \\ \hline
$17$ & \multicolumn{1}{|l|}{$\pi^{D/2}(-m^{2})^{\sigma}(z_3)^{-c}(z_4)^{-a}\frac{(-a)_{c}}{(a-\sigma)_{c+D/2}}$} \\ 
\hline
$18$ & \multicolumn{1}{|l|}{$\pi^{D/2}(-m^{2})^{\sigma}(z_3)^{-b}(z_4)^{-a}\frac{(-a)_{b}}{(a-\sigma))_{b+D/2}}$} \\ 
\hline
$19$ & \multicolumn{1}{|l|}{$\pi^{D/2}(-m^{2})^{\sigma}(-z_1)^{-b}(z_4)^{-a}\frac{(-b)_{a}}{(b-\sigma)_{a+D/2}}$} \\ \hline
$20$ & \multicolumn{1}{|l|}{$\pi^{D/2}(-m^{2})^{\sigma}(-z_1)^{-c}(z_4)^{-a}\frac{(-c)_{a}}{(c-\sigma)_{a+D/2}}$} \\ \hline
$21$ & \multicolumn{1}{|l|}{$\pi^{D/2}(-m^{2})^{\sigma}(-z_3)^{-c}(-z_4)^{-a}\frac{1}{(-b-D/2)_{D/2}}$} \\ \hline
$22$ & \multicolumn{1}{|l|}{$\pi^{D/2}(-m^{2})^{\sigma}(-z_3)^{-b}(-z_4)^{-a}\frac{1}{(-c-D/2)_{D/2}}$} \\ \hline
\end{tabular}
\ 
\]
\[
{\ 
\begin{tabular}{|l|c|}
\hline
$n$ & $C_{n}$ \\ \hline\hline
$23$ & \multicolumn{1}{|l|}{$\pi^{D/2}(-m^{2})^{\sigma}(z_4)^{c-\sigma}\frac{(-a)_{2a+D/2}(-b)_{2b+D/2}}{(c-\sigma)_{-2c+2\sigma+D/2}}$} \\ \hline
$24$ & \multicolumn{1}{|l|}{$\pi^{D/2}(-m^{2})^{\sigma}(z_4)^{b-\sigma}\frac{(-a)_{2a+D/2}(-c)_{2c+D/2}}{(b-\sigma)_{-2b+2\sigma+D/2}}$} \\ \hline
$25$ & \multicolumn{1}{|l|}{$\pi^{D/2}(-m^{2})^{\sigma}(z_2)^{-b}\frac{(-a)_{b}(-c)_{-b+\sigma}}{(b-\sigma)_{\sigma+D/2}}$} \\ \hline
$26$ & \multicolumn{1}{|l|}{$\pi^{D/2}(-m^{2})^{\sigma}(z_2)^{-c}\frac{(-a)_{c}(-b)_{-c+\sigma}}{(c-\sigma)_{\sigma+D/2}}$} \\ \hline
$27$ & \multicolumn{1}{|l|}{$\pi^{D/2}(-m^{2})^{\sigma}(-z_1)^{-c}\frac{(-b)_{-a-D/2}}{(-b+\sigma)_{-a}}$} \\ \hline
$28$ & \multicolumn{1}{|l|}{$\pi^{D/2}(-m^{2})^{\sigma}(-z_1)^{-b}\frac{(-c)_{-a-D/2}}{(-c+\sigma)_{-a}}$} \\ \hline
$29$ & \multicolumn{1}{|l|}{$\pi^{D/2}(-m^{2})^{\sigma}(-z_1)^{a+D/2}\frac{(-b)_{-a-D/2}(-c)_{-a-D/2}}{(a+D/2)_{-2a-D/2}}$} \\ \hline \hline
$30$ & \multicolumn{1}{|l|}{$\pi^{D/2}(-m^{2})^{\sigma}(-z_3)^{-a}\frac{(-b)_{a}}{(a-\sigma)_{a+D/2}}$} \\ \hline
$31$ & \multicolumn{1}{|l|}{$\pi^{D/2}(-m^{2})^{\sigma}(-z_3)^{-a}\frac{(-c)_{a}}{(a-\sigma)_{a+D/2}}$} \\ \hline
$32$ & \multicolumn{1}{|l|}{$\pi^{D/2}(-m^{2})^{\sigma}\frac{(D/2)_{a}}{(-\sigma)_{a+D/2}}$} \\ \hline
\multicolumn{2}{c}{Table-5}
\end{tabular}
\ }
\]
\[
\begin{tabular}{|c|c|c|}
\hline
$n$ & \multicolumn{1}{|c|}{$x_{1},\,x_{2},\,x_{3};\,x_{4},\,x_{5}$} & $z_{1};z_{2};z_{3};z_{4}$ \\ \hline\hline
$1$ & $-\sigma,\,-c,\,1-\sigma-D/2;\,1+a-\sigma,\,1+b-\sigma $ & $-\frac{m^{2}}{p^{2}};\,\frac{q^{2}}{p^{2}};\,\frac{r^{2}}{p^{2}};\,\frac{m^{2}}{p^{2}}$ \\ \hline
$2$ & $-\sigma,\,-b,\,1-\sigma-D/2;\,1+a-\sigma,\,1+c-\sigma $ & $-\frac{m^{2}}{q^{2}};\,\frac{p^{2}}{q^{2}};\,\frac{r^{2}}{q^{2}};\,\frac{m^{2}}{q^{2}}$ \\ \hline
$3$ & $-\sigma,\,-a,\,1-\sigma-D/2;\,1+c-\sigma,\,1+b-\sigma $ & $\frac{p^{2}}{r^{2}};\,\frac{q^{2}}{r^{2}};\,\frac{m^{2}}{r^{2}};\,\frac{m^{2}}{r^{2}}$ \\ \hline
$4$ & $-c,\,b+D/2,\,1-\sigma-D/2;\,1-c+\sigma,\,1+b-\sigma $ & $\frac{q^{2}}{r^{2}};\,\frac{p^{2}}{r^{2}};\,\frac{m^{2}}{r^{2}};\,\frac{m^{2}}{p^{2}}$ \\ \hline
$5$ & $-b,\,c+D/2,\,1-\sigma-D/2;\,1-b+\sigma,\,1+c-\sigma $ & $\frac{p^{2}}{r^{2}};\,\frac{q^{2}}{r^{2}};\,\frac{m^{2}}{r^{2}};\,\frac{m^{2}}{q^{2}}$ \\ \hline
$6$ & \multicolumn{1}{|c|}{$-c,\,a+D/2,\,1-\sigma-D/2;\,1+a-\sigma,\,1-c+\sigma $} & \multicolumn{1}{|c|}{$\frac{r^{2}}{q^{2}};\,\frac{m^{2}}{q^{2}};\,\frac{p^{2}}{q^{2}};\,-\frac{m^{2}}{p^{2}}$} \\ \hline
$7$ & \multicolumn{1}{|c|}{$-b,\,a+D/2,\,1-\sigma-D/2;\,1+a-\sigma,\,1-b+\sigma $} & \multicolumn{1}{|c|}{$\frac{r^{2}}{p^{2}};\,\frac{m^{2}}{p^{2}};\,\frac{q^{2}}{p^{2}};\,-\frac{m^{2}}{q^{2}}$} \\ \hline
$8$ & $c+D/2,\,-a,\,1-\sigma-D/2;\,1-a+\sigma,\,1+c-\sigma $ & $\frac{r^{2}}{q^{2}};\,\frac{p^{2}}{q^{2}};\,\frac{m^{2}}{r^{2}};\,-\frac{m^{2}}{r^{2}}$ \\ \hline
$9$ & $b+D/2,\,-a,\,1-\sigma-D/2;\,1-a+\sigma,\,1+b-\sigma $ & $\frac{r^{2}}{p^{2}};\,\frac{q^{2}}{p^{2}};\,\frac{m^{2}}{r^{2}};\,-\frac{m^{2}}{r^{2}}$ \\ \hline
$10$ & \multicolumn{1}{|c|}{$\sigma+D/2,\,c+D/2,\,1-\sigma-D/2;\,1-b+\sigma,\,1-a+\sigma$} & \multicolumn{1}{|c|}{$\frac{r^{2}}{p^{2}};\,\frac{q^{2}}{p^{2}};\,-\frac{m^{2}p^{2}}{q^{2}r^{2}};\,\frac{m^{2}}{r^{2}}$} \\ \hline
$11$ & \multicolumn{1}{|c|}{$\sigma+D/2,\,b+D/2,\,1-\sigma-D/2;\,1-c+\sigma,\,1-a+\sigma$} & \multicolumn{1}{|c|}{$\frac{r^{2}}{q^{2}};\,\frac{p^{2}}{q^{2}};\,-\frac{m^{2}q^{2}}{p^{2}r^{2}};\,\frac{m^{2}}{r^{2}}$} \\ \hline
$12$ & \multicolumn{1}{|c|}{$\sigma+D/2,\,a+D/2,\,1-\sigma-D/2;\,1-b+\sigma,\,1-c+\sigma$} & \multicolumn{1}{|c|}{$\frac{q^{2}}{r^{2}};\,\frac{p^{2}}{r^{2}};\,\frac{m^{2}}{q^{2}};\,\frac{m^{2}}{p^{2}}$} \\ \hline
$13$ & \multicolumn{1}{|c|}{$c+D/2,\,a+D/2,\,1-a-D/2;\,1-a+\sigma,\,1-b+\sigma $} & \multicolumn{1}{|c|}{$\frac{q^{2}}{p^{2}};\,\frac{m^{2}}{p^{2}};\,-\frac{p^{2}}{q^{2}};\,\frac{r^{2}}{m^{2}}$} \\ \hline
$14$ & \multicolumn{1}{|c|}{$b+D/2,\,a+D/2,\,1-a-D/2;\,1-a+\sigma,\,1-c+\sigma $} & \multicolumn{1}{|c|}{$\frac{p^{2}}{q^{2}};\,\frac{m^{2}}{q^{2}};\,-\frac{q^{2}}{p^{2}};\,\frac{r^{2}}{m^{2}}$} \\ \hline
$15$ & \multicolumn{1}{|c|}{$b+D/2,\,a+D/2,\,b-\sigma;\,b+D/2,\,1-c+\sigma $} & \multicolumn{1}{|c|}{$\frac{p^{2}}{r^{2}};\,\frac{q^{2}}{m^{2}};\,-\frac{m^{2}}{p^{2}};\,\frac{m^{2}}{r^{2}}$} \\ \hline
$16$ & \multicolumn{1}{|c|}{$c+D/2,\,a+D/2,\,c-\sigma;\,c+D/2,1-b+\sigma $} & \multicolumn{1}{|c|}{$\frac{q^{2}}{r^{2}};\,\frac{p^{2}}{m^{2}};\,-\frac{m^{2}}{q^{2}};\,\frac{m^{2}}{r^{2}}$} \\ \hline
$17$ & \multicolumn{1}{|c|}{$-c,\,a-\sigma,\,a+D/2;\,a+D/2,\,1+a-c$} & 
\multicolumn{1}{|c|}{$\frac{p^{2}}{q^{2}};\,-\frac{p^{2}r^{2}}{m^{2}q^{2}};\,\frac{p^{2}}{q^{2}};\,-\frac{m^{2}}{p^{2}}$} \\ \hline
$18$ & \multicolumn{1}{|c|}{$-b,\,a-\sigma,\,a+D/2;\,a+D/2,\,1+a-b$} & 
\multicolumn{1}{|c|}{$\frac{q^{2}}{p^{2}};\,-\frac{q^{2}r^{2}}{m^{2}p^{2}};\,\frac{q^{2}}{p^{2}};\,-\frac{m^{2}}{q^{2}}$} \\ \hline
$19$ & \multicolumn{1}{|c|}{$a-b,\,1-b-D/2,\,-a;\,1-b+\sigma,\,1-b-D/2$} &
\multicolumn{1}{|c|}{$\frac{m^{2}}{r^{2}};\,\frac{m^{2}}{r^{2}};\,-\frac{q^{2}r^{2}}{m^{2}p^{2}};\,\frac{r^{2}}{p^{2}}$} \\ \hline
$20$ & \multicolumn{1}{|c|}{$a-c,\,1-c-D/2,\,-a;\,1-c+\sigma,\,1-c-D/2$} &\multicolumn{1}{|c|}{$\frac{m^{2}}{r^{2}};\,\frac{m^{2}}{r^{2}};\,-\frac{p^{2}r^{2}}{m^{2}q^{2}};\,\frac{r^{2}}{q^{2}}$} \\ \hline
$21$ & $1+b-\sigma,\,-c,\,-a;\,1+b+D/2,\,1+b-\sigma $ & $-\frac{m^{2}}{q^{2}}{p^{2}r^{2}};\,\frac{m^{2}}{r^{2}};\,\frac{m^{2}}{r^{2}};\,\frac{m^{2}}{p^{2}}$ \\ 
\hline
$22$ & $1+c-\sigma,\,-a,\,-b;\,1+c+D/2,\,1+c-\sigma $ & $-\frac{m^{2}p^{2}}{q^{2}r^{2}};\,\frac{m^{2}}{r^{2}};\,\frac{m^{2}}{r^{2}};\,\frac{m^{2}}{q^{2}}$ \\ \hline
\end{tabular}
\ 
\]

\bigskip

\[
{\ 
\begin{tabular}{|c|l|l|}
\hline
$n$ & \multicolumn{1}{|c|}{$x_{1},\,x_{2},\,x_{3};\,x_{4},\,x_{5}$} & $z_{1};z_{2};z_{3};z_{4}$ \\ \hline\hline
$23$ & \multicolumn{1}{|l|}{$-c,\,b+D/2,\,a+D/2;\,1-c+\sigma,\,a+b+D$} & 
\multicolumn{1}{|c|}{$\frac{q^{2}}{m^{2}};\,-\frac{p^{2}}{m^{2}};\,\frac{r^{2}}{m^{2}};\,-\frac{m^{2}}{p^{2}}$} \\ \hline
$24$ & \multicolumn{1}{|l|}{$-b,\,c+D/2,\,a+D/2;\,1-b+\sigma,\,a+c+D$} & 
\multicolumn{1}{|c|}{$\frac{p^{2}}{m^{2}};\,-\frac{q^{2}}{m^{2}};\,\frac{r^{2}}{m^{2}};\,-\frac{m^{2}}{q^{2}}$} \\ \hline
$25$ & \multicolumn{1}{|l|}{$-b,\,1-b-D/2,\,a+D/2;\,1-b+\sigma,\,1+a-b$} & \multicolumn{1}{|c|}{$\frac{m^{2}}{p^{2}};\,-\frac{m^{2}}{p^{2}};\,\frac{r^{2}}{p^{2}};\,-\frac{q^{2}}{m^{2}}$} \\ \hline
$26$ & \multicolumn{1}{|l|}{$-c,\,1-c-D/2,\,a+D/2;\,1-c+\sigma,\,1+a-c$} & 
\multicolumn{1}{|c|}{$\frac{m^{2}}{q^{2}};\,-\frac{m^{2}}{q^{2}};\,\frac{r^{2}}{q^{2}};\,-\frac{p^{2}}{m^{2}}$} \\ \hline
$27$ & $-c,\,1-c-D/2,\,-a;\,1-c+\sigma,\,1+b-\sigma $ & $\frac{m^{2}}{r^{2}};\,\frac{m^{2}}{r^{2}};\,\frac{q^{2}}{r^{2}};\,\frac{p^{2}}{m^{2}}$ \\ \hline
$28$ & $-b,\,1-b-D/2,\,-a;\,1-b+\sigma,\,1+c-\sigma $ & $\frac{m^{2}}{r^{2}};\,\frac{m^{2}}{r^{2}};\,\frac{p^{2}}{r^{2}};\,\frac{q^{2}}{m^{2}}$ \\ \hline
$29$ & \multicolumn{1}{|l|}{$1+a,\,a+D/2,\,-a;\,1-c+\sigma,\,1-b+\sigma$} & 
\multicolumn{1}{|c|}{$\frac{m^{2}}{r^{2}};\,\frac{m^{2}}{r^{2}};\,\frac{p^{2}}{m^{2}};\,\frac{q^{2}}{m^{2}}$} \\ \hline \hline
$n $ & \multicolumn{1}{|c|}{$x_1,\,x_2,\,x_3,\,x_4,\,x_5;\,x_6,\,x_7$} & \multicolumn{1}{|c|}{$z_1,\,z_2,\,z_3$} \\ \hline\hline
$30$ & $-c,\,-a,\,1\!-\!a\!-\!D/2,\,1\!-\!a\!+\!b\!+\!c,\,a\!+\!D/2;\,1\!-\!a\!+\!b,\,1\!-\!a\!+\!\sigma $ & $-\frac{r^{2}}{m^{2}};\,\frac{q^{2}}{p^{2}};\,\frac{m^{2}}{p^{2}}$ \\ \hline
$31$ & $-b,\,-a,\,1\!-\!a\!-\!D/2,\,1\!-\!a\!+\!b\!+\!c,\,a\!+\!D/2;\,1\!-\!a\!+\!c,\,1\!-\!a\!+\!\sigma $ & $-\frac{r^{2}}{m^{2}};\,\frac{p^{2}}{q^{2}};\,\frac{m^{2}}{q^{2}}$ \\ \hline
$32$ & $-\sigma,\,-a,\,-b,\,-c,\,a+D/2;\,D/2,\,-b-c$ & $\frac{p^{2}}{m^{2}};\,\frac{q^{2}}{m^{2}};\,\frac{r^{2}}{m^{2}}$ \\ \hline
\multicolumn{2}{c}{Table-6} 
\end{tabular}
} 
\]

\bigskip
\bigskip

\section{Three massive denominators}

$ $

\bigskip

The three massive denominators solutions $J_{n}=J_{n}(a,b,c,D,p,q,m,m,m),$
where $n=1,2,...,40$, are given by hypergeometric functions listed below. 
\bigskip

\begin{eqnarray*}
\Psi _{1}\left[ 
\begin{array}{l|}
x_{1},x_{2},x_{3} \\ 
x_{4},x_{5}
\end{array}
\;\,z_{1};z_{2};z_{3};z_{4};z_{5}\right]  &=&{\ \sum_{j_{1},..,j_{5}=0}^{\infty
}}\frac{
(x_{1})_{j_{1}+j_{2}+j_{3}}(x_{2})_{j_{1}+j_{2}-j_{4}}(x_{3})_{j_{1}+j_{3}-j_{5}}
}{(x_{4})_{j_{1}-j_{4}-j_{5}}(x_{5})_{j_{1}+j_{2}+j_{3}-j_{4}-j_{5}}} \\
&&\times \frac{z_{1}^{j_{1}}}{j_{1}!}\frac{z_{2}^{j_{2}}}{j_{2}!}\frac{
z_{3}^{j_{3}}}{j_{3}!}\frac{z_{4}^{j_{4}}}{j_{4}!}\frac{z_{5}^{j_{5}}}{j_{5}!
}, \\
\Psi _{2}\left[ 
\begin{array}{l|}
x_{1},x_{2},x_{3} \\ 
x_{4},x_{5}
\end{array}
\;\,z_{1};z_{2};z_{3};z_{4};z_{5}\right]  &=&{\ \sum_{j_{1},..,j_{5}=0}^{\infty
}}\frac{
(x_{1})_{j_{1}+j_{2}+j_{5}}(x_{2})_{j_{3}+j_{4}+j_{5}}(x_{3})_{j_{1}+j_{2}-j_{4}}
}{(x_{4})_{j_{2}-j_{3}-j_{4}}
(x_{5})_{j_{1}+j_{3}+j_{5}}} \\
&&\times \frac{z_{1}^{j_{1}}}{j_{1}!}\frac{z_{2}^{j_{2}}}{j_{2}!}\frac{
z_{3}^{j_{3}}}{j_{3}!}\frac{z_{4}^{j_{4}}}{j_{4}!}\frac{z_{5}^{j_{5}}}{j_{5}!
}, \\
\Psi _{3}\left[ 
\begin{array}{l|}
x_{1},x_{2},x_{3} \\ 
x_{4},x_{5}
\end{array}
\;\,z_{1};z_{2};z_{3};z_{4};z_{5}\right]  &=&{\ \sum_{j_{1},..,j_{5}=0}^{\infty
}}\frac{
(x_{1})_{j_{1}+j_{2}+j_{3}+j_{4}+j_{5}}(x_{2})_{j_{1}+j_{2}+j_{4}}(x_{3})_{j_{1}+j_{3}+j_{5}}
}{(x_{4})_{j_{1}+j_{4}+j_{5}}(x_{5})_{j_{1}+j_{2}+j_{3}}} \\
&&\times \frac{z_{1}^{j_{1}}}{j_{1}!}\frac{z_{2}^{j_{2}}}{j_{2}!}\frac{
z_{3}^{j_{3}}}{j_{3}!}\frac{z_{4}^{j_{4}}}{j_{4}!}\frac{z_{5}^{j_{5}}}{j_{5}!
}, \\
\Psi _{4}\left[ 
\begin{array}{l|}
x_{1},x_{2},x_{3} \\ 
x_{4},x_{5}
\end{array}
\;\,z_{1};z_{2};z_{3};z_{4};z_{5}\right]  &=&{\sum_{j_{1},..,j_{5}=0}^{\infty }}
\frac{
(x_{1})_{j_{1}+j_{2}-j_{3}}(x_{2})_{j_{1}+j_{4}-j_{5}}(x_{3})_{-j_{2}+j_{3}+j_{4}}
}{(x_{4})_{j_{1}-j_{3}-j_{5}}
(x_{5})_{-j_{2}+j_{4}-j_{5}}} \\
&&\times \frac{z_{1}^{j_{1}}}{j_{1}!}\frac{
z_{2}^{j_{2}}}{j_{2}!}\frac{z_{3}^{j_{3}}}{j_{3}!}\frac{z_{4}^{j_{4}}}{j_{4}!
}\frac{z_{5}^{j_{5}}}{j_{5}!}, \\
R_{10}\left[ 
\begin{array}{l|}
x_{1},x_{2},x_{3},x_{4} \\ 
\multicolumn{1}{c|}{x_{5},x_{6},x_{7}}
\end{array}
\;\,z_{1};z_{2};z_{3};z_{4}\right]  &=&{\sum_{j_{1},..,j_{4}=0}^{\infty }}\frac{
(x_{1})_{j_{1}+j_{2}+j_{3}}(x_{2})_{j_{1}+j_{2}-j_{4}}(x_{3})_{j_{1}+j_{2}-2j_{4}}(x_{4})_{j_{3}+j_{4}}
}{(x_{5})_{j_{1}+j_{2}+j_{3}-j_{4}}(x_{6})_{j_{2}-j_{4}}(x_{7})_{j_{1}-j_{4}}
} \\
&&\times \frac{z_{1}^{j_{1}}{}}{j_{1}!}\frac{z_{2}^{j_{2}}}{j_{2}!}\frac{
z_{3}^{j_{3}}}{j_{3}!}\frac{z_{4}^{j_{4}}}{j_{4}!}, \\
T_{11}\left[ 
\begin{array}{l|}
x_{1},x_{2},x_{3},x_{4} \\ 
\multicolumn{1}{c|}{x_{5}}
\end{array}
\;\,z_{1};z_{2};z_{3}\right]  &=&{\sum_{j_{1},j_{2},j_{3}=0}^{\infty }}\frac{
(x_{1})_{j_{1}+j_{2}+j_{3}}(x_{2})_{j_{1}+j_{2}}(x_{3})_{j_{1}+j_{3}}}{
(x_{5})_{2j_{1}+2j_{2}+2j_{3}}} \\
&&\times (x_{4})_{j_{2}+j_{3}}\frac{z_{1}^{j_{1}}}{j_{1}!}\frac{z_{2}^{j_{2}}
}{j_{2}!}\frac{z_{3}^{j_{3}}}{j_{3}!}
\end{eqnarray*}
and the expression of each one solution is given by $J_{n}=D_{n}\Psi _{l},$
where the relation between $n$ and $l$ os given by 
\[
\begin{tabular}{|c|c|c|c|c|}
\hline
$n$ & $1,...,15$ & $16,...,21$ & $22,...,27$ & $28,...,33$ \\ \hline
$l$ & $1$ & $2$ & $3$ & $4$ \\ \hline
\end{tabular}
\ ,
\]
or $J_{n}=D_{n}R_{10},$ $n=34,...,39;$ $J_{40}=D_{40}T_{11},$ where
the
 coefficients $D_{n}$ are shown in Table-7, the parameters and
variables of
 the functions $\Psi _{l},R_{10}$, and $T_{11}$ in the
Table-8 below. 
\[
\begin{tabular}{|c|c|}
\hline
$n$ & $D_{n}$ \\ \hline\hline
$1$ & \multicolumn{1}{|l|}{$\pi^{D/2}(r^{2})^{\sigma}(-z_4)^{\sigma-c}(-z_5)^{\sigma-b}\frac{(-b)_{-a-D/2}(-c)_{-a-D/2}}{(a+D/2)_{-2a-D/2}}$} \\ \hline\hline
$2$ & \multicolumn{1}{|l|}{$\pi^{D/2}(q^{2})^{\sigma}(-z_4)^{\sigma-a}(-z_5)^{\sigma-c}\frac{(-a)_{-b-D/2}(-c)_{-b-D/2}}{(b+D/2)_{-2b-D/2}}$} \\ \hline\hline
$3$ & \multicolumn{1}{|l|}{$\pi^{D/2}(p^{2})^{\sigma}(-z_4)^{\sigma-b}(-z_5)^{\sigma-a}\frac{(-a)_{-c-D/2}(-b)_{-c-D/2}}{(c+D/2)_{-2c-D/2}}$} \\ \hline\hline
$4$ & \multicolumn{1}{|l|}{$\pi^{D/2}(-m^{2})^{\sigma}(z_3)^{-c}(z_4)^{-a}\frac{(-b)_{-c-D/2}}{(-a+c)_{-c}}$} \\ \hline\hline
$5$ & \multicolumn{1}{|l|}{$\pi^{D/2}(-m^{2})^{\sigma}(z_3)^{-b}(z_4)^{-a}\frac{(-c)_{-b-D/2}}{(-a+b)_{-b}}$} \\ \hline\hline
$6$ & \multicolumn{1}{|l|}{$\pi^{D/2}(-m^{2})^{\sigma}(z_3)^{-c}(z_4)^{-b}\frac{(-a)_{-c-D/2}}{(-b+c)_{-c}}$} \\ \hline\hline
$7$ & \multicolumn{1}{|l|}{$\pi^{D/2}(-m^{2})^{\sigma}(z_3)^{-a}(z_4)^{-b}\frac{(-c)_{-a-D/2}}{(-b+a)_{-a}}$} \\ \hline\hline
$8$ & \multicolumn{1}{|l|}{$\pi^{D/2}(-m^{2})^{\sigma}(z_3)^{-b}(z_4)^{-c}\frac{(-a)_{-b-D/2}}{(-c+b)_{-b}}$} \\ \hline\hline
$9$ & \multicolumn{1}{|l|}{$\pi^{D/2}(-m^{2})^{\sigma}(z_3)^{-a}(z_4)^{-c}\frac{(-b)_{-a-D/2}}{(-c+a)_{-a}}$} \\ \hline\hline
$10$ & \multicolumn{1}{|l|}{$\pi^{D/2}(-m^{2})^{\sigma}(z_5)^{c+D/2}\frac{(-a)_{-c-D/2}(-b)_{-c-D/2}}{(c+D/2)_{-2c-D/2}}$} \\ \hline
$11$ & \multicolumn{1}{|l|}{$\pi^{D/2}(-m^{2})^{\sigma}(z_5)^{c-\sigma}\frac{(-a)_{2a+D/2}(-b)_{2b+D/2}}{(c-\sigma)_{-2c+2\sigma+D/2}}$} \\ \hline\hline
$12$ & \multicolumn{1}{|l|}{$\pi^{D/2}(-m^{2})^{\sigma}(z_5)^{b+D/2}\frac{(-a)_{-b-D/2}(-c)_{-b-D/2}}{(b+D/2)_{-2b-D/2}}$} \\ \hline
$13$ & \multicolumn{1}{|l|}{$\pi^{D/2}(-m^{2})^{\sigma}(z_5)^{b-\sigma}\frac{(-a)_{2a+D/2}(-c)_{2c+D/2}}{(b-\sigma)_{-2b+2\sigma+D/2}}$} \\ \hline\hline
$14$ & \multicolumn{1}{|l|}{$\pi^{D/2}(-m^{2})^{\sigma}(z_5)^{a-\sigma}\frac{(-b)_{2b+D/2}(-c)_{2c+D/2}}{(a-\sigma)_{-2a+2\sigma+D/2}}$} \\ \hline
$15$ & \multicolumn{1}{|l|}{$\pi^{D/2}(-m^{2})^{\sigma}(z_5)^{a+D/2}\frac{(-b)_{-a-D/2}(-c)_{-a-D/2}}{(a+D/2)_{-2a-D/2}}$} \\ \hline\hline
$16$ & \multicolumn{1}{|l|}{$\pi^{D/2}(p^{2})^{\sigma}(z_4)^{-c}\frac{(-a)_{\sigma}(-b)_{-c+\sigma}}{(c-\sigma)_{-c+2\sigma+D/2}}$} \\ \hline\hline
$17$ & \multicolumn{1}{|l|}{$\pi^{D/2}(q^{2})^{\sigma}(z_4)^{-b}\frac{(-a)_{\sigma}(-c)_{-b+\sigma}}{(b-\sigma)_{-b+2\sigma+D/2}}$} \\ \hline \hline
$18$ & \multicolumn{1}{|l|}{$\pi^{D/2}(p^{2})^{\sigma}(z_4)^{-c}\frac{(-a)_{-c+\sigma}(-b)_{\sigma}}{(c-\sigma)_{-c+2\sigma+D/2}}$} \\ \hline\hline
$19$ & \multicolumn{1}{|l|}{$\pi^{D/2}(r^{2})^{\sigma}(z_4)^{-a}\frac{(-b)_{\sigma}(-c)_{-a+\sigma}}{(a-\sigma)_{-a+2\sigma+D/2}}$} \\ \hline\hline
$20$ & \multicolumn{1}{|l|}{$\pi^{D/2}(q^{2})^{\sigma}(z_4)^{-b}\frac{(-a)_{-b+\sigma}(-c)_{\sigma}}{(b-\sigma)_{-b+2\sigma+D/2}}$} \\ \hline\hline$21$ & \multicolumn{1}{|l|}{$\pi^{D/2}(r^{2})^{\sigma}(z_4)^{-a}\frac{(-b)_{-a+\sigma}(-c)_{\sigma}}{(a-\sigma)_{-a+2\sigma+D/2}}$} \\ \hline\hline
$22$ & \multicolumn{1}{|l|}{$\pi^{D/2}(-m^{2})^{\sigma}(-z_4)^{-b}(-z_5)^{-c}\frac{1}{(-a-D/2)_{D/2}}$} \\ \hline\hline
$23$ & \multicolumn{1}{|l|}{$\pi^{D/2}(-m^{2})^{\sigma}(-z_4)^{-a}(-z_5)^{-c}\frac{1}{(-b-D/2)_{D/2}}$} \\ \hline\hline
$24$ & \multicolumn{1}{|l|}{$\pi^{D/2}(-m^{2})^{\sigma}(-z_4)^{-a}(-z_5)^{-b}\frac{1}{(-c-D/2)_{D/2}}$} \\ \hline\hline

\end{tabular}
\ 
\]
\[
\begin{tabular}{|c|c|}
\hline
$n$ & $D_{n}$ \\ \hline\hline
$25$ & \multicolumn{1}{|l|}{$\pi^{D/2}(p^{2})^{\sigma}\frac{(-a)_{\sigma}(-b)_{\sigma}}{(-\sigma)_{2\sigma+D/2}}$} \\ \hline\hline
$26$ & \multicolumn{1}{|l|}{$\pi^{D/2}(q^{2})^{\sigma}\frac{(-a)_{\sigma}(-c)_{\sigma}}{(-\sigma)_{2\sigma+D/2}}$} \\ \hline\hline
$27$ & \multicolumn{1}{|l|}{$\pi^{D/2}(r^{2})^{\sigma}\frac{(-b)_{\sigma}(-c)_{\sigma}}{(-\sigma)_{2\sigma+D/2}}$} \\ \hline\hline

$28$ & \multicolumn{1}{|l|}{$\pi^{D/2}(-m^{2})^{\sigma}(z_1)^{c+D/2}(-z_3)^{-a}\frac{(-a)_{-c-D/2}(-b)_{-c-D/2}}{(c+D/2)_{-2c-D/2}}$} \\ \hline
$29$ & \multicolumn{1}{|l|}{$\pi^{D/2}(-m^{2})^{\sigma}(z_1)^{b+D/2}(-z_3)^{-a}\frac{(-a)_{-b-D/2}(-c)_{-b-D/2}}{(b+D/2)_{-2b-D/2}}$} \\ \hline\hline
$30$ & \multicolumn{1}{|l|}{$\pi^{D/2}(-m^{2})^{\sigma}(z_4)^{c+D/2}(z_5)^{-b}\frac{(-a)_{-c-D/2}(-b)_{-c-D/2}}{(c+D/2)_{-2c-D/2}}$} \\ \hline

$31$ & \multicolumn{1}{|l|}{$\pi^{D/2}(-m^{2})^{\sigma}(z_4)^{a+D/2}(z_5)^{-b}\frac{(-b)_{-a-D/2}(-c)_{-a-D/2}}{(a+D/2)_{-2a-D/2}}$} \\ \hline\hline
$32$ & \multicolumn{1}{|l|}{$\pi^{D/2}(-m^{2})^{\sigma}(z_4)^{b+D/2}(z_5)^{-c}\frac{(-a)_{-b-D/2}(-c)_{-b-D/2}}{(b+D/2)_{-2b-D/2}}$} \\ \hline
$33$ & \multicolumn{1}{|l|}{$\pi^{D/2}(-m^{2})^{\sigma}(z_4)^{a+D/2}(z_5)^{-c}\frac{(-b)_{-a-D/2}(-c)_{-a-D/2}}{(a+D/2)_{-2a-D/2}}$} \\ \hline\hline
$34$ & \multicolumn{1}{|l|}{$\pi^{D/2}(-m^{2})^{\sigma}(-z_1)^{-b}\frac{(-a)_{-c-D/2}}{(-a+b)_{-c}}$} \\ \hline\hline
$35$ & \multicolumn{1}{|l|}{$\pi^{D/2}(-m^{2})^{\sigma}(-z_1)^{-a}\frac{(-b)_{-c-D/2}}{(a-b)_{-c}}$} \\ \hline\hline
$36$ & \multicolumn{1}{|l|}{$\pi^{D/2}(-m^{2})^{\sigma}(-z_1)^{-c}\frac{(-a)_{-b-D/2}}{(-a+c)_{-b}}$} \\ \hline\hline

$37$ & \multicolumn{1}{|l|}{$\pi^{D/2}(-m^{2})^{\sigma}(-z_1)^{-a}\frac{(-c)_{-b-D/2}}{(a-c)_{-b}}$} \\ \hline \hline

$38$ & \multicolumn{1}{|l|}{$\pi^{D/2}(-m^{2})^{\sigma}(-z_1)^{-c}\frac{(-b)_{-a-D/2}}{(-b+c)_{-a}}$} \\ \hline\hline

$39$ & \multicolumn{1}{|l|}{$\pi^{D/2}(-m^{2})^{\sigma}(-z_1)^{-b}\frac{(-c)_{-a-D/2}}{(b-c)_{-a}}$} \\ \hline\hline

$40$ & \multicolumn{1}{|l|}{$\pi^{D/2}(-m^{2})^{\sigma}\frac{1}{(-\sigma)_{D/2}}$}
\\ \hline
\multicolumn{2}{c}{Table-7}
\end{tabular}
\ 
\]

\[
\begin{tabular}{|c|c|c|}
\hline
$n$ & \multicolumn{1}{|c|}{$x_{1},\,x_{2},\,x_{3};\,x_{4},\,x_{5}$} & $
z_{1};z_{2};z_{3};z_{4};z_{5}$ \\ \hline\hline
$1$ & \multicolumn{1}{|l|}{$1-\sigma-D/2,\,c-\sigma,\,b-\sigma;\,1-a-D/2,\,1-\sigma-D/2$} & \multicolumn{1}{|l|}{$-\frac{m^{2}r^{2}}{p^{2}q^{2}};\,\frac{m^{2}}{p^{2}};\,\frac{m^{2}}{q^{2}};\,-\frac{p^{2}}{r^{2}};\,-\frac{q^{2}}{r^{2}}$} \\ \hline\hline
$2$ & \multicolumn{1}{|l|}{$1-\sigma-D/2,\,a-\sigma,\,c-\sigma;\,1-b-D/2,\,1-\sigma-D/2$} & \multicolumn{1}{|l|}{$-\frac{m^{2}q^{2}}{p^{2}r^{2}};\,\frac{m^{2}}{r^{2}};\,\frac{m^{2}}{p^{2}};\,-\frac{r^{2}}{q^{2}};\,-\frac{p^{2}}{q^{2}}$} \\ \hline\hline
$3$ & \multicolumn{1}{|l|}{$1-\sigma-D/2,\,b-\sigma,\,a-\sigma;\,1-c-D/2,\,1-\sigma-D/2$} & \multicolumn{1}{|l|}{$-\frac{m^{2}p^{2}}{q^{2}r^{2}};\,\frac{m^{2}}{q^{2}};\,\frac{m^{2}}{r^{2}};\,-\frac{q^{2}}{p^{2}};\,-\frac{r^{2}}{p^{2}}$} \\ \hline\hline
$4$ & \multicolumn{1}{|l|}{$-c,\,a+D/2,\,a-\sigma;\,a+D/2,\,1+a-c$} & 
\multicolumn{1}{|l|}{$-\frac{p^{2}r^{2}}{m^{2}q^{2}};\,\frac{p^{2}}{q^{2}};\,\frac{p^{2}}{q^{2}};\,-\frac{m^{2}}{p^{2}};\,-\frac{m^{2}}{p^{2}}$} \\
\hline\hline
$5$ & \multicolumn{1}{|l|}{$-b,\,a+D/2,\,a-\sigma;\,a+D/2,\,1+a-b$} & 
\multicolumn{1}{|l|}{$-\frac{q^{2}r^{2}}{m^{2}p^{2}};\,\frac{q^{2}}{p^{2}};\,\frac{q^{2}}{p^{2}};\,-\frac{m^{2}}{q^{2}};\,-\frac{m^{2}}{q^{2}}$} \\ 
\hline\hline
$6$ & \multicolumn{1}{|l|}{$-c,\,b-\sigma,\,b+D/2;\,b+D/2,\,1-c+b$} & 
\multicolumn{1}{|l|}{$-\frac{p^{2}q^{2}}{m^{2}r^{2}};\,\frac{p^{2}}{r^{2}};\,\frac{p^{2}}{r^{2}};\,-\frac{m^{2}}{p^{2}};\,-\frac{m^{2}}{p^{2}}$} \\
\hline\hline
$7$ & \multicolumn{1}{|l|}{$-a,\,b-\sigma,\,b+D/2;\,b+D/2,\,1-a+b$} & 
\multicolumn{1}{|l|}{$-\frac{q^{2}r^{2}}{m^{2}p^{2}};\,\frac{r^{2}}{p^{2}};\,\frac{r^{2}}{p^{2}};\,-\frac{m^{2}}{r^{2}};\,-\frac{m^{2}}{r^{2}}$} \\ 
\hline\hline
$8$ & \multicolumn{1}{|l|}{$-b,\,c-\sigma,\,c+D/2;\,c+D/2,\,1-b+c$} & 
\multicolumn{1}{|l|}{$-\frac{p^{2}q^{2}}{m^{2}r^{2}};\,\frac{q^{2}}{r^{2}};\,\frac{q^{2}}{r^{2}};\,-\frac{m^{2}}{q^{2}};\,-\frac{m^{2}}{q^{2}}$} \\ 
\hline\hline
$9$ & \multicolumn{1}{|l|}{$-a,\,c-\sigma,\,c+D/2;\,c+D/2,\,1-a+c$} & 
\multicolumn{1}{|l|}{$-\frac{p^{2}r^{2}}{m^{2}q^{2}};\,\frac{r^{2}}{q^{2}};\,\frac{r^{2}}{q^{2}};\,-\frac{m^{2}}{r^{2}};\,-\frac{m^{2}}{r^{2}}$} \\ 
\hline\hline
$10$ & \multicolumn{1}{|l|}{$-c,\,b-\sigma,\,a-\sigma;\,1-c-D/2,\,-c$} & 
\multicolumn{1}{|l|}{$-\frac{p^{2}}{m^{2}};\,\frac{q^{2}}{m^{2}};\,\frac{r^{2}}{m^{2}};\,-\frac{m^{2}}{p^{2}};\,-\frac{m^{2}}{p^{2}}$} \\ \hline
$11$ & \multicolumn{1}{|l|}{$-c,\,b+D/2,\,a+D/2,\,1-c+\sigma,\,a+b+D$} & 
\multicolumn{1}{|l|}{$-\frac{p^{2}}{m^{2}};\,\frac{q^{2}}{m^{2}};\,\frac{r^{2}}{m^{2}};\,-\frac{m^{2}}{p^{2}};\,-\frac{m^{2}}{p^{2}}$} \\ \hline
\hline
$12$ & \multicolumn{1}{|l|}{$-b,\,c-\sigma,\,a-\sigma;\,1-b-D/2,\,-b$} & 
\multicolumn{1}{|l|}{$-\frac{q^{2}}{m^{2}};\,\frac{p^{2}}{m^{2}};\,\frac{r^{2}}{m^{2}};\,-\frac{m^{2}}{q^{2}};-\frac{m^{2}}{q^{2}}$} \\ \hline
$13$ & \multicolumn{1}{|l|}{$-b,\,c+D/2,\,a+D/2;\,1-b+\sigma,\,a+c+D$} & 
\multicolumn{1}{|l|}{$-\frac{q^{2}}{m^{2}};\,\frac{p^{2}}{m^{2}};\,\frac{r^{2}}{m^{2}};\,-\frac{m^{2}}{q^{2}};\,-\frac{m^{2}}{q^{2}}$} \\ \hline\hline
$14$ & \multicolumn{1}{|l|}{$-a,\,c+D/2,\,b+D/2;\,1-a+\sigma,\,b+c+D$} & 
\multicolumn{1}{|l|}{$-\frac{r^{2}}{m^{2}};\,\frac{p^{2}}{m^{2}};\,\frac{q^{2}}{m^{2}};\,-\frac{m^{2}}{r^{2}};\,-\frac{m^{2}}{r^{2}}$} \\ \hline
$15$ & \multicolumn{1}{|l|}{$-a,\,c-\sigma,\,b-\sigma;\,1-a-D/2,\,-a$} & 
\multicolumn{1}{|l|}{$-\frac{r^{2}}{m^{2}};\,\frac{p^{2}}{m^{2}};\,\frac{q^{2}}{m^{2}};\,-\frac{m^{2}}{r^{2}};\,-\frac{m^{2}}{r^{2}}$} \\ \hline\hline
$16$ & \multicolumn{1}{|l|}{$1-\sigma-D/2,\,a+D/2,\,-c;\,1-c+\sigma,\,1+a-\sigma$} & \multicolumn{1}{|l|}{$\frac{m^{2}}{p^{2}};\,\frac{m^{2}}{p^{2}};\,-\frac{r^{2}}{q^{2}};\,\frac{p^{2}}{q^{2}};\,\frac{m^{2}}{q^{2}}$} \\ \hline\hline
$17$ & \multicolumn{1}{|l|}{$1-\sigma-D/2,\,-b,\,b-\sigma;\,1-a-D/2,1+a-\sigma $} & \multicolumn{1}{|l|}{$\frac{m^{2}}{q^{2}};\,\frac{m^{2}}{q^{2}};\,-\frac{r^{2}}{p^{2}};\,\frac{q^{2}}{p^{2}};\,\frac{m^{2}}{p^{2}}$} \\ \hline \hline
$18$ & \multicolumn{1}{|l|}{$1-\sigma-D/2,\,-c,\,c-\sigma;\,1-b-D/2,\,1+b-\sigma $} & \multicolumn{1}{|l|}{$\frac{m^{2}}{p^{2}};\,\frac{m^{2}}{p^{2}};\,-\frac{q^{2}}{r^{2}};\,\frac{p^{2}}{r^{2}};\,\frac{m^{2}}{r^{2}}$} \\ \hline\hline
$19$ & \multicolumn{1}{|l|}{$1-\sigma-D/2,\,-a,\,a-\sigma;\,1-b-D/2,\,1+b-\sigma $} & \multicolumn{1}{|l|}{$\frac{m^{2}}{r^{2}};\,\frac{m^{2}}{r^{2}};\,-\frac{q^{2}}{p^{2}};\,\frac{r^{2}}{p^{2}};\,\frac{m^{2}}{p^{2}}$} \\ \hline\hline
$20$ & \multicolumn{1}{|l|}{$1-\sigma-D/2,\,-b,\,b-\sigma;\,1-c-D/2,\,1+c-\sigma $} & \multicolumn{1}{|l|}{$\frac{m^{2}}{q^{2}};\,\frac{m^{2}}{q^{2}};\,-\frac{p^{2}}{r^{2}};\,\frac{q^{2}}{r^{2}};\,\frac{m^{2}}{r^{2}}$} \\ \hline\hline
$21$ & \multicolumn{1}{|l|}{$1-\sigma -D/2,-a,a-\sigma,1-c-D/2,1-\sigma +c$}
& \multicolumn{1}{|l|}{$\frac{m^{2}}{r^{2}};\,\frac{m^{2}}{r^{2}};\,-\frac{p^{2}}{q^{2}};\,\frac{r^{2}}{q^{2}};\,\frac{m^{2}}{q^{2}}$} \\ \hline\hline
$22$ & \multicolumn{1}{|l|}{$1+a-\sigma,\,-b,\,-c;\,1+a+D/2,\,1+a-\sigma$} & \multicolumn{1}{|l|}{$-\frac{m^{2}r^{2}}{p^{2}q^{2}};\,\frac{m^{2}}{p^{2}};\,\frac{m^{2}}{q^{2}};\,\frac{m^{2}}{p^{2}};\,\frac{m^{2}}{q^{2}}$} \\ \hline\hline
$23$ & \multicolumn{1}{|l|}{$1+b-\sigma,\,-a,\,-c;\,1+b+D/2,\,1+b-\sigma$} & \multicolumn{1}{|l|}{$-\frac{m^{2}q^{2}}{p^{2}r^{2}};\,\frac{m^{2}}{p^{2}};\,\frac{m^{2}}{r^{2}};\,\frac{m^{2}}{p^{2}};\,\frac{m^{2}}{r^{2}}$} \\ \hline\hline
$24$ & \multicolumn{1}{|l|}{$1+c-\sigma,\,-a,\,-b;\,1+c+D/2,\,1+c-\sigma$} & \multicolumn{1}{|l|}{$\!\!-\frac{m^{2}p^{2}}{q^{2}r^{2}};\,\frac{m^{2}}{q^{2}};\,\frac{m^{2}}{r^{2}};\,\frac{m^{2}}{q^{2}};\,\frac{m^{2}}{r^{2}}$} \\ \hline\hline

\end{tabular}
\ 
\]
\[
\begin{tabular}{|c|c|c|}
\hline
$n$ & \multicolumn{1}{|c|}{$x_{1},\,x_{2},\,x_{3};\,x_{4},\,x_{5}$} & $
z_{1};z_{2};z_{3};z_{4};z_{5}$ \\ \hline\hline

$25$ & \multicolumn{1}{|l|}{$-\sigma,\,-c,1-\sigma-D/2;\,1+a-\sigma,\,1+b-\sigma $} & \multicolumn{1}{|l|}{$-\frac{m^{2}}{p^{2}};\,\frac{q^{2}}{p^{2}};\,\frac{m^{2}}{p^{2}};\,\frac{r^{2}}{p^{2}};\,\frac{m^{2}}{p^{2}}$} \\ \hline\hline
$26$ & \multicolumn{1}{|l|}{$-\sigma,\,-b,\,1-\sigma-D/2;\,1+a-\sigma,\,1+c-\sigma $} & \multicolumn{1}{|l|}{$\!\!-\frac{m^{2}}{q^{2}};\,\frac{p^{2}}{q^{2}};\,\frac{m^{2}}{q^{2}};\,\frac{r^{2}}{q^{2}};\,\frac{m^{2}}{q^{2}}$} \\ \hline\hline
$27$ & \multicolumn{1}{|l|}{$-\sigma,\,-a,\,1-\sigma-D/2;\,1+b-\sigma,\,1+c-\sigma $} & \multicolumn{1}{|l|}{$\!\!-\frac{m^{2}}{r^{2}};\,\frac{p^{2}}{r^{2}};\,\frac{m^{2}}{r^{2}};\,\frac{q^{2}}{r^{2}};\,\frac{m^{2}}{r^{2}}$} \\ \hline\hline
$28$ & \multicolumn{1}{|l|}{$a-\sigma,a+D/2,\,c+D/2;\,a+D/2;\,1-b+\sigma$} & \multicolumn{1}{|l|}{$\frac{r^{2}}{m^{2}};\,\frac{p^{2}}{q^{2}};\,\frac{m^{2}}{p^{2}};\,\frac{q^{2}}{p^{2}};\,-\frac{p^{2}}{q^{2}}$} \\ \hline
$29$ & \multicolumn{1}{|l|}{$a-\sigma,a+D/2,\,b+D/2,\,a+D/2;\,1-c+\sigma$} & \multicolumn{1}{|l|}{$\frac{r^{2}}{m^{2}};\,\frac{q^{2}}{p^{2}};\,\frac{m^{2}}{q^{2}};\,\frac{p^{2}}{q^{2}};\,-\frac{m^{2}}{p^{2}}$} \\ \hline\hline
$30$ & \multicolumn{1}{|l|}{$b-\sigma,\,b+D/2,\,c+D/2;\,b+D/2,\,1-a+\sigma$} & \multicolumn{1}{|l|}{$\frac{q^{2}}{m^{2}};\,\frac{p^{2}}{r^{2}};\,\frac{m^{2}}{p^{2}};\,\frac{r^{2}}{p^{2}};\,-\frac{m^{2}}{r^{2}}$} \\ \hline
$31$ & \multicolumn{1}{|l|}{$b-\sigma,\,b+D/2,\,a+D/2;\,b+D/2,\,1-c+\sigma$} & \multicolumn{1}{|l|}{$\frac{q^{2}}{m^{2}};\,\frac{r^{2}}{p^{2}};\,\frac{m^{2}}{r^{2}};\,\frac{p^{2}}{r^{2}};\,-\frac{m^{2}}{p^{2}}$} \\ \hline\hline
$32$ & \multicolumn{1}{|l|}{$c-\sigma,\,c+D/2,\,b+D/2;\,c+D/2,\,1-a+\sigma$} & \multicolumn{1}{|l|}{$\frac{p^{2}}{m^{2}};\,\frac{q^{2}}{r^{2}};\,\frac{m^{2}}{q^{2}};\,\frac{r^{2}}{q^{2}};\,-\frac{m^{2}}{r^{2}}$} \\ \hline
$33$ & \multicolumn{1}{|l|}{$c-\sigma,\,c+D/2,\,a+D/2;\,c+D/2,\,1-b+\sigma$} & \multicolumn{1}{|l|}{$\frac{p^{2}}{m^{2}};\,\frac{r^{2}}{q^{2}};\,\frac{m^{2}}{r^{2}};\,\frac{q^{2}}{r^{2}};\,-\frac{m^{2}}{q^{2}}$} \\ \hline\hline
$n$ & \multicolumn{1}{|c|}{$x_{1},\,x_{2},\,x_{3},\,x_{4};\,x_{5},\,x_{6},\,x_{7}$} & $z_{1};\,z_{2};\,z_{3};\,z_{4}$ \\ \hline\hline
$34$ & \multicolumn{1}{|l|}{$\!\!-b,\,1\!-\!b\!-\!D/2,\,1\!+\!a\!-\!b\!+\!c,\,-c;\,1\!+\!a\!-\!b,\,1\!-\!b\!-\!D/2,\,1\!-\!b\!+\!\sigma\!\!\!\!\! $ } & \multicolumn{1}{|c|}{$\frac{m^{2}}{p^{2}};\,\frac{m^{2}}{p^{2}};\,\frac{r^{2}}{p^{2}};\,\frac{q^{2}}{m^{2}}$} \\ \hline\hline
$35$ & \multicolumn{1}{|l|}{$\!\!-a,\,1\!-\!a\!-\!D/2,\,1\!-\!a\!+\!b\!+\!c,\,-c;\,1\!-\!a\!+\!b,\,1\!-\!a\!-\!D/2,\,1\!-\!a\!+\!\sigma $ } & \multicolumn{1}{|c|}{$\frac{m^{2}}{p^{2}};\,\frac{m^{2}}{p^{2}};\,\frac{q^{2}}{p^{2}};\,\frac{r^{2}}{m^{2}}$} \\ \hline\hline
$36$ & \multicolumn{1}{|l|}{$\!\!-c,\,1\!-\!c\!-\!D/2,\,1\!+\!a\!+\!b\!-\!c,\,-b;\,1\!+\!a\!-\!c,\,1\!-\!c\!-\!D/2,\,1\!-\!c\!+\!\sigma $ } & \multicolumn{1}{|c|}{$\frac{m^{2}}{q^{2}};\,\frac{m^{2}}{q^{2}};\,\frac{r^{2}}{q^{2}};\,\frac{p^{2}}{m^{2}}$} \\ \hline\hline

$37$ & \multicolumn{1}{|l|}{$\!\!-a,\,1\!-\!a\!-\!D/2,\,1\!-\!a\!+\!b\!+\!c,\,-b;\,1\!-\!a\!+\!c,\,1\!-\!a\!-\!D/2,\,1\!-\!a\!+\!\sigma $ } & \multicolumn{1}{|c|}{$\frac{m^{2}}{q^{2}};\,\frac{m^{2}}{q^{2}};\,\frac{p^{2}}{q^{2}};\,\frac{r^{2}}{m^{2}}$} \\ \hline \hline

$38$ & \multicolumn{1}{|l|}{$\!\!-c,\,1\!-\!c\!-\!D/2,\,1\!+\!a\!+\!b\!-\!c,\,-a;\,1\!+\!b\!-\!c,\,1\!-\!c\!-\!D/2,\,1\!-\!c\!+\!\sigma $ } & \multicolumn{1}{|c|}{$\frac{m^{2}}{r^{2}};\,\frac{m^{2}}{r^{2}};\,\frac{q^{2}}{r^{2}};\,\frac{p^{2}}{m^{2}}$} \\ \hline\hline

$39$ & \multicolumn{1}{|l|}{$\!\!-b,\,1\!-\!b\!-\!D/2,\,1\!+\!a\!-\!b\!+\!c,\,-a;\,1\!-\!b\!+\!c,\,1\!-\!b\!-\!D/2,\,1\!-\!b\!+\!\sigma $ } & \multicolumn{1}{|c|}{$\frac{m^{2}}{r^{2}};\,\frac{m^{2}}{r^{2}};\,\frac{p^{2}}{r^{2}};\,\frac{q^{2}}{m^{2}}$} \\ \hline\hline

$n$ & \multicolumn{1}{|c|}{$x_{1},\,x_{2},\,x_{3},\,x_{4};\,x_{5} $} & $ z_{1};\,z_{2};\,z_{3}$ \\ \hline\hline
$40$ & \multicolumn{1}{|l|}{$-\sigma,\,-a,\,-b,\,-c;\,-\sigma+D/2$} & 
\multicolumn{1}{|c|}{$\frac{p^{2}}{m^{2}};\,\frac{q^{2}}{m^{2}};\,\frac{r^{2}}{
m^{2}}$} \\ \hline
\multicolumn{2}{c}{Table-8} 

\end{tabular}
\ 
\]
\[
\begin{tabular}{|c|c|c|}
\hline

\end{tabular}
\ 
\]

\section{Partial differential equations satisfied by the hypergeometric-type functions}
$ $

\bigskip

Some of the functions presented in this paper can be expressed in
terms of the generalized hyergeometric functions of Lauricella,
namely, the functions 
$
F_{4},T_{1},T_{7},T_{8},T_{10},T_{11},R_{6}$ and $R_{9}.$ The others
are hypergeometric-type functions that satisfies one of the sets with 
respect to the partial differential equations, given below.

In writing the partial differential equations we use the operator 
\begin{equation}
d_{n}=z_{n}\frac{\partial }{\partial z_{n}},
\end{equation}
where $n=1,2,...,7.$

$T_{2}$ satisfies the equations 
\begin{eqnarray*}
\{(d_{1}+d_{2}-d_{3}+x_{1})(d_{1}+d_{2}-d_{3}+x_{2})-\frac{1}{z_{1}}%
d_{1}(d_{1}-d_{3}+x_{4}-1)\}F &=&0, \\
\{(d_{1}+d_{2}-d_{3}+x_{1})(d_{1}+d_{2}-d_{3}+x_{2})-\frac{1}{z_{2}}%
d_{2}(d_{2}-d_{3}+x_{5}-1)\}F &=&0, \\
\{(d_{3}+x_{3})-\frac{1}{z_{3}}%
d_{3}(d_{1}-d_{3}+x_{4}-1)(d_{2}-d_{3}+x_{5}-1)\}F &=&0,
\end{eqnarray*}
where $F=F(z_{1},z_{2},z_{3})$ but, as in cases below, $F$ can be function 
of $z_{4}$,and or $z_{5}$ too.

$T_{3}$ satisfies the equations 
\begin{eqnarray*}
\{(d_{1}+d_{2}-d_{3}+x_{1})(d_{1}+d_{2}+x_{2})-\frac{1}{z_{1}}%
d_{1}(d_{1}+x_{5}-1)\}F &=&0, \\
\{(d_{1}+d_{2}-d_{3}+x_{1})(d_{1}+d_{2}+x_{2})-\frac{1}{z_{2}}%
d_{2}(d_{2}-d_{3}+x_{4}-1)\}F &=&0, \\
\{(d_{1}+d_{2}-d_{3}+x_{1})(d_{3}+x_{3})-\frac{1}{z_{3}}%
d_{3}(d_{2}-d_{3}+x_{4}-1)\}F &=&0.
\end{eqnarray*}

$T_{4}$ satisfies the equations 
\begin{eqnarray*}
\{(d_{1}+d_{2}+x_{1})(d_{1}+d_{2}+x_{2})(d_{1}-d_{3}+x_{3})\\
-\frac{1}{z_{1}}d_{1}(d_{1}-d_{3}+x_{4}-1)(d_{1}+d_{2}-d_{3}+x_{5}-1)\}F &=&0, \\
\{(d_{1}+d_{2}+x_{1})(d_{1}+d_{2}+x_{2})-\frac{1}{z_{2}}%
d_{2}(d_{1}+d_{2}-d_{3}+x_{5}-1)\}F &=&0, \\
\{(d_{1}-d_{3}+x_{3})
-\frac{1}{z_{3}}d_{3}(d_{1}-d_{3}+x_{4}-1)(d_{1}+d_{2}-d_{3}+x_{5}-1)\}F &=&0.
\end{eqnarray*}

$T_{5}$ satisfies the equations 
\begin{eqnarray*}
\{(d_{1}+d_{2}+x_{1})(d_{1}+d_{3}+x_{2})-\frac{1}{z_{1}}d_{1}(d_{1}+x_{5}-1)\}F &=&0, \\
\{(d_{1}+d_{2}+x_{1})(d_{2}-d_{3}+x_{3})-\frac{1}{z_{2}}%
d_{2}(d_{2}-d_{3}+x_{4}-1)\}F &=&0, \\
\{(d_{1}+d_{3}+x_{2})(d_{2}-d_{3}+x_{3})-\frac{1}{z_{3}}%
d_{3}(d_{2}-d_{3}+x_{4}-1)\}F &=&0.
\end{eqnarray*}

$T_{6}$ satisfies the equations 
\begin{eqnarray*}
\{(d_{1}+d_{2}+x_{1})(d_{1}+d_{2}+x_{2})-\frac{1}{z_{1}}%
d_{1}(d_{1}+d_{3}+x_{5}-1)\}F &=&0, \\
\{(d_{1}+d_{2}+x_{1})(d_{1}+d_{2}+x_{2})-\frac{1}{z_{2}}%
d_{2}(d_{2}-d_{3}+x_{4}-1)\}F &=&0, \\
\{(d_{3}+x_{3})-\frac{1}{z_{3}}%
d_{3}(d_{2}-d_{3}+x_{4}-1)(d_{1}+d_{3}+x_{5}-1)\}F &=&0.
\end{eqnarray*}

$T_{9}$ satisfies the equations 
\begin{eqnarray*}
\{(d_{1}+d_{2}+x_{1})(-d_{1}+d_{3}+x_{3})(-2d_{1}+d_{3}+x_{4})(d_{1}+x_{5})\\-
\frac{1}{z_{1}}d_{1}(-d_{1}+d_{2}+d_{3}+x_{6}-1)(-d_{1}+d_{3}+x_{7}-1)\}F
&=&0, \\
\{(d_{1}+d_{2}+x_{1})(d_{2}+d_{3}+x_{2})-\frac{1}{z_{2}}%
d_{2}(-d_{1}+d_{2}+d_{3}+x_{6}-1)\}F &=&0, \\
\{(d_{2}+d_{3}+x_{2})(-d_{1}+d_{3}+x_{3})(-2d_{1}+d_{3}+x_{4})\\ -\frac{1}{z_{3}%
}d_{3}(-d_{1}+d_{2}+d_{3}+x_{6}-1)(-d_{1}+d_{3}+x_{7}-1)\}F &=&0.
\end{eqnarray*}

$R_{1}$ satisfies the equations
\begin{eqnarray*}
\{(d_{1}+d_{2}+d_{3}+x_{1})(d_{1}+d_{2}-d_{4}+x_{2})-\frac{1}{z_{1}}%
d_{1}(d_{1}+d_{2}+d_{3}-d_{4}+x_{5}-1)\}F &=&0, \\
\{(d_{1}+d_{2}+d_{3}+x_{1})(d_{1}+d_{2}-d_{4}+x_{2})(d_{2}+d_{3}+x_{3})\\
-\frac{1}{z_{2}}d_{2}(d_{2}-d_{4}+x_{4}-1)(d_{1}+d_{2}+d_{3}-d_{4}+x_{5}-1)%
\}F &=&0, \\
\{(d_{1}+d_{2}+d_{3}+x_{1})(d_{2}+d_{3}+x_{3})-\frac{1}{z_{3}}%
d_{3}(d_{1}+d_{2}+d_{3}-d_{4}+x_{5}-1)\}F &=&0, \\
\{(d_{1}+d_{2}-d_{4}+x_{2})-\frac{1}{z_{4}}%
d_{4}(d_{2}-d_{4}+x_{4}-1)(d_{1}+d_{2}+d_{3}-d_{4}+x_{5}-1)\}F &=&0.
\end{eqnarray*}

$R_{2}$ satisfies the equations
\begin{eqnarray*}
\{(d_{1}+d_{2}-d_{3}-d_{4}+x_{1})(d_{1}+d_{2}+x_{2})-\frac{1}{z_{1}}%
d_{1}(d_{1}-d_{3}+x_{4}-1)\}F &=&0, \\
\{(d_{1}+d_{2}-d_{3}-d_{4}+x_{1})(d_{1}+d_{2}+x_{2})-\frac{1}{z_{2}}%
d_{2}(d_{2}-d_{4}+x_{5}-1)\}F &=&0, \\
\{(d_{1}+d_{2}-d_{3}-d_{4}+x_{1})(d_{3}+d_{4}+x_{4})-\frac{1}{z_{3}}%
d_{3}(d_{1}-d_{3}+x_{4}-1)\}F &=&0, \\
\{(d_{1}+d_{2}-d_{3}-d_{4}+x_{1})(d_{3}+d_{4}+x_{4})-\frac{1}{z_{4}}%
d_{42}(d_{2}-d_{4}+x_{5}-1)\}F &=&0.
\end{eqnarray*}

$R_{3}$ satisfies the equations
\begin{eqnarray*}
\{(d_{1}+d_{2}-d_{3}-d_{4}+x_{1})(d_{1}+d_{2}-d_{3}+x_{2})-\frac{1}{z_{1}}%
d_{1}(d_{1}-d_{3}-d_{4}+x_{5}-1)\}F &=&0, \\
\{(d_{1}+d_{2}-d_{3}-d_{4}+x_{1})(d_{1}+d_{2}-d_{3}+x_{2})-\frac{1}{z_{2}}%
d_{2}(d_{2}-d_{3}+x_{4}-1)\}F &=&0, \\
\{(d_{1}+d_{2}-d_{3}-d_{4}+x_{1})(d_{1}+d_{2}-d_{3}+x_{2})(d_{3}+d_{4}+x_{3})\\
-\frac{1}{z_{3}}d_{3}(d_{2}-d_{3}+x_{4}-1)(d_{1}-d_{3}-d_{4}+x_{5}-1)\}F &=&0,
\\
\{(d_{1}+d_{2}-d_{3}-d_{4}+x_{1})(d_{3}+d_{4}+x_{3})-\frac{1}{z_{4}}%
d_{4}(d_{1}-d_{3}-d_{4}+x_{5}-1)\}F &=&0.
\end{eqnarray*}

$R_{4}$ satisfies the equations
\begin{eqnarray*}
\{(d_{1}+d_{2}-d_{3}+x_{1})(d_{1}+d_{4}+x_{2})-\frac{1}{z_{1}}%
d_{1}(d_{1}-d_{3}+x_{5}-1)\}F &=&0, \\
\{(d_{1}+d_{2}-d_{3}+x_{1})(d_{2}-d_{4}+x_{3})-\frac{1}{z_{2}}%
d_{2}(d_{2}-d_{3}-d_{4}+x_{4}-1)\}F &=&0, \\
\{(d_{1}+d_{2}-d_{3}+x_{1})-\frac{1}{z_{3}}%
d_{3}(d_{2}-d_{3}-d_{4}+x_{4}-1)(d_{1}-d_{3}+x_{5}-1)\}F &=&0, \\
\{(d_{1}+d_{4}+x_{2})(d_{2}-d_{4}+x_{3})-\frac{1}{z_{4}}%
d_{4}(d_{2}-d_{3}-d_{4}+x_{4}-1)\}F &=&0.
\end{eqnarray*}

$R_{5}$ satisfies the equations
\begin{eqnarray*}
\{(d_{1}+d_{2}+d_{3}+x_{1})(d_{1}+d_{3}+x_{2})-\frac{1}{z_{1}}%
d_{1}(d_{1}+d_{2}+d_{4}+x_{4}-1)\}F &=&0, \\
\{(d_{1}+d_{2}+d_{3}+x_{1})(d_{2}+d_{4}+x_{3})-\frac{1}{z_{2}}%
d_{2}(d_{1}+d_{2}+d_{4}+x_{4}-1)\}F &=&0, \\
\{(d_{1}+d_{2}+d_{3}+x_{1})(d_{1}+d_{3}+x_{2})-\frac{1}{z_{3}}%
d_{3}(d_{3}-d_{4}+x_{5}-1)\}F &=&0, \\
\{(d_{2}+d_{4}+x_{3})-\frac{1}{z_{4}}%
d_{4}(d_{1}+d_{2}+d_{4}+x_{4}-1)(d_{3}-d_{4}+x_{5}-1)\}F &=&0.
\end{eqnarray*}

$R_{7}$ satisfies the equations
\begin{eqnarray*}
\{(d_{1}+d_{2}+d_{3}+x_{1})(d_{1}+d_{2}-d_{4}+x_{2})-\frac{1}{z_{1}}%
d_{1}(d_{1}+d_{3}+x_{5}-1)\}F &=&0, \\
\{(d_{1}+d_{2}+d_{3}+x_{1})(d_{1}+d_{2}-d_{4}+x_{2})-\frac{1}{z_{2}}%
d_{2}(d_{2}-d_{4}+x_{4}-1)\}F &=&0, \\
\{(d_{1}+d_{2}+d_{3}+x_{1})(d_{3}+d_{4}+x_{3})-\frac{1}{z_{3}}%
d_{3}(d_{1}+d_{3}+x_{5}-1)\}F &=&0, \\
\{(d_{1}+d_{2}-d_{4}+x_{2})(d_{3}+d_{4}+x_{3})-\frac{1}{z_{4}}%
d_{4}(d_{2}-d_{4}+x_{4}-1)\}F &=&0.
\end{eqnarray*}

$R_{8}$ satisfies the equations
\begin{eqnarray*}
\{(d_{1}+d_{2}-d_{3}+x_{1})(d_{1}+d_{2}+x_{2})-\frac{1}{z_{1}}%
d_{1}(d_{1}-d_{3}-d_{4}+x_{4}-1)\}F &=&0, \\
\{(d_{1}+d_{2}-d_{3}+x_{1})(d_{1}+d_{2}+x_{2})-\frac{1}{z_{2}}%
d_{2}(d_{2}+d_{4}+x_{5}-1)\}F &=&0, \\
\{(d_{1}+d_{2}-d_{3}+x_{1})(d_{3}+d_{4}+x_{3})-\frac{1}{z_{3}}%
d_{3}(d_{1}-d_{3}-d_{4}+x_{4}-1)\}F &=&0, \\
\{(d_{3}+d_{4}+x_{3})-\frac{1}{z_{4}}%
d_{4}(d_{1}-d_{3}-d_{4}+x_{4}-1)(d_{2}+d_{4}+x_{5}-1)\}F &=&0.
\end{eqnarray*}

$R_{10}$ satisfies the equations
\begin{eqnarray*}
\{(d_{1}+d_{2}+d_{3}+x_{1})(d_{1}+d_{2}-d_{4}+x_{2})(d_{1}+d_{2}-2d_{4}+x_{3})\\
-\frac{1}{z_{1}}d_{1}(d_{1}+d_{2}+d_{3}-d_{4}+x_{5}-1)(d_{1}-d_{4}+x_{7}-1)%
\}F &=&0, \\
\{(d_{1}+d_{2}+d_{3}+x_{1})(d_{1}+d_{2}-d_{4}+x_{2})(d_{1}+d_{2}-2d_{4}+x_{3})\\
-\frac{1}{z_{2}}d_{2}(d_{1}+d_{2}+d_{3}-d_{4}+x_{5}-1)(d_{2}-d_{4}+x_{6}-1)%
\}F &=&0, \\
\{(d_{1}+d_{2}+d_{3}+x_{1})(d_{3}+d_{4}+x_{4})-\frac{1}{z_{3}}%
d_{3}(d_{1}+d_{2}+d_{3}-d_{4}+x_{5}-1)\}F &=&0, \\
\{(d_{1}+d_{2}-d_{4}+x_{2})(d_{1}+d_{2}-2d_{4}+x_{3})(d_{3}+d_{4}+x_{4})-%
\frac{1}{z_{4}}%
\end{eqnarray*}

$\Psi _{1}$ satisfies the equations
\begin{eqnarray*}
\{(d_{1}+d_{2}+d_{3}+x_{1})(d_{1}+d_{2}-d_{4}+x_{2})(d_{1}+d_{3}-d_{5}+x_{3})\\
-\frac{1}{z_{1}}%
d_{1}(d_{1}-d_{4}-d_{5}+x_{4}-1)(d_{1}+d_{2}+d_{3}-d_{4}-d_{5}+x_{5}-1)\}F
&=&0, \\
\{(d_{1}+d_{2}+d_{3}+x_{1})(d_{1}+d_{2}-d_{4}+x_{2})-\frac{1}{z_{2}}%
d_{2}(d_{1}+d_{2}+d_{3}-d_{4}-d_{5}+x_{5}-1)\}F &=&0, \\
\{(d_{1}+d_{2}+d_{3}+x_{1})(d_{1}+d_{3}-d_{5}+x_{3})-\frac{1}{z_{3}}%
d_{3}(d_{1}+d_{2}+d_{3}-d_{4}-d_{5}+x_{5}-1)\}F &=&0, \\
\{(d_{1}+d_{2}-d_{4}+x_{2})-\frac{1}{z_{4}}%
d_{4}(d_{1}-d_{4}-d_{5}+x_{4}-1)(d_{1}+d_{2}+d_{3}-d_{4}-d_{5}+x_{5}-1)\}F
&=&0, \\
\{(d_{1}+d_{3}-d_{5}+x_{3})-\frac{1}{z_{5}}%
d_{5}(d_{1}-d_{4}-d_{5}+x_{4}-1)(d_{1}+d_{2}+d_{3}-d_{4}-d_{5}+x_{5}-1)\}F
&=&0.
\end{eqnarray*}

$\Psi _{2}$ satisfies the equations
\begin{eqnarray*}
\{(d_{1}+d_{2}+d_{5}+x_{1})(d_{1}+d_{2}-d_{4}+x_{3})-\frac{1}{z_{1}}%
d_{1}(d_{1}+d_{3}+d_{5}+x_{5}-1)\}F &=&0, \\
\{(d_{1}+d_{2}+d_{5}+x_{1})(d_{1}+d_{2}-d_{4}+x_{3})-\frac{1}{z_{2}}%
d_{2}(d_{2}-d_{3}-d_{4}+x_{4}-1)\}F &=&0, \\
\{(d_{3}+d_{4}+d_{5}+x_{2})-\frac{1}{z_{3}}%
d_{3}(d_{2}-d_{3}-d_{4}+x_{4}-1)(d_{1}+d_{3}+d_{5}+x_{5}-1)\}F &=&0, \\
\{(d_{3}+d_{4}+d_{5}+x_{2})(d_{1}+d_{2}-d_{4}+x_{3})-\frac{1}{z_{4}}%
d_{4}(d_{2}-d_{3}-d_{4}+x_{4}-1)\}F &=&0, \\
\{(d_{1}+d_{2}+d_{5}+x_{1})(d_{3}+d_{4}+d_{5}+x_{2})-\frac{1}{z_{5}}%
d_{5}(d_{1}+d_{3}+d_{5}+x_{5}-1)\}F &=&0.
\end{eqnarray*}

$\Psi _{3}$ satisfies the equations 
\begin{eqnarray*}
\{(d_{1}+d_{2}+d_{3}+d_{4}+d_{5}+x_{1})(d_{1}+d_{2}+d_{4}+x_{2})(d_{1}+d_{3}+d_{5}+x_{3})\\-\frac{1}{z_{1}}d_{1}(d_{1}+d_{4}+d_{5}+x_{4}-1)(d_{1}+d_{2}+d_{3}+x_{5}-1)%
\}F &=&0, \\
\{(d_{1}+d_{2}+d_{3}+d_{4}+d_{5}+x_{1})(d_{1}+d_{2}+d_{4}+x_{2})-\frac{1}{%
z_{2}}d_{2}(d_{1}+d_{2}+d_{3}+x_{5}-1)\}F &=&0, \\
\{(d_{1}+d_{2}+d_{3}+d_{4}+d_{5}+x_{1})(d_{1}+d_{3}+d_{5}+x_{3})-\frac{1}{%
z_{3}}d_{3}(d_{1}+d_{2}+d_{3}+x_{5}-1)\}F &=&0, \\
\{(d_{1}+d_{2}+d_{3}+d_{4}+d_{5}+x_{1})(d_{1}+d_{2}+d_{4}+x_{2})-\frac{1}{%
z_{4}}d_{4}(d_{1}+d_{4}+d_{5}+x_{4}-1)\}F &=&0, \\
\{(d_{1}+d_{2}+d_{3}+d_{4}+d_{5}+x_{1})(d_{1}+d_{3}+d_{5}+x_{3})-\frac{1}{%
z_{5}}d_{5}(d_{1}+d_{4}+d_{5}+x_{4}-1)\}F &=&0.
\end{eqnarray*}

$\Psi _{4}$ satisfies the equations
\begin{eqnarray*}
\{(d_{1}+d_{2}-d_{3}+x_{1})(d_{1}+d_{4}-d_{5}+x_{2})-\frac{1}{z_{1}}%
d_{1}(d_{1}-d_{3}-d_{5}+x_{4}-1)\}F &=&0, \\
\{(d_{1}+d_{2}-d_{3}+x_{1})(-d_{2}+d_{3}+d_{4}+x_{3})-\frac{1}{z_{2}}%
d_{2}(-d_{2}+d_{4}-d_{5}+x_{5}-1)\}F &=&0, \\
\{(d_{1}+d_{2}-d_{3}+x_{1})(-d_{2}+d_{3}+d_{4}+x_{3})-\frac{1}{z_{3}}%
d_{3}(d_{1}-d_{3}-d_{5}+x_{4}-1)\}F &=&0, \\
\{(d_{1}+d_{4}-d_{5}+x_{2})(-d_{2}+d_{3}+d_{4}+x_{3})-\frac{1}{z_{4}}%
d_{4}(-d_{2}+d_{4}-d_{5}+x_{5}-1)\}F &=&0, \\
\{(d_{1}+d_{4}-d_{5}+x_{2})-\frac{1}{z_{5}}%
d_{5}(d_{1}-d_{3}-d_{5}+x_{4}-1)(-d_{2}+d_{4}-d_{5}+x_{5}-1)\}F &=&0.
\end{eqnarray*}

\bigskip


\bigskip

\bigskip

\bigskip

\bigskip

\begin{thebibliography}{99}
\bibitem{hoof/Veltman} G. \ `t Hooft and M. Veltman, Nucl. Phys. B 44 (1972)
189.

\bibitem{giambiagi} C. G. Bollini and J. J. Giambiagi, Nuovo Cimento B, 12
(1972) 20.

\bibitem{davydy} \'{E}. \'{E}. Boos and A. I. Davydychev, Theor. Math. Phys.
89 (1991) 1052.

\bibitem{terrano} A. E. Terrano, Phys. Lett. B 93 (1980) 424.

\bibitem{tkachov} F. V. Tkachov, Phys. Lett. B 100 (1981) 65.

\bibitem{smirnov} V. A. Smirnov, Nucl. Phys. B 566 (2000) 469; Phys. Lett. B
460 (1999) 397.

\bibitem{laporta} S. Laporta, E. Remiddi, Phys. Lett. B 356 (1995) 390;
Phys. Lett. B 379 (1996) 283. V. W. Hughes, T. Kinoshita, Rev. Mod. Phys. 71
(1999) S133. S Laporta, E. Remiddi, Acta Phys. Pol. B 28 (1997) 959.

\bibitem{Anastasiou} C. Anastasiou, E. W. N. Glover, C. Oleari, Nucl. Phys.
B 575 (2000) 416. Erratum-ibid B 585 (2000) 763.

\bibitem{bern} Z. Bern, L. Dixon, D. A. Kosower, JHEP 1 (2000) 27.

\bibitem{gehrmann} T. Gehrmann, E. Remiddi, Nucl. Phys. B 601 (2001) 287;
Nucl. Phys. Proc. Suppl. 89 (2000) 251.

\bibitem{chetyrkin} K. Chetyrkin, M. Misiak, M. M\"unz, Nucl. Phys. B 518
(1998) 473.

\bibitem{glover} E. W. N. Glover, J. B. Tausk, J. J. van der Bij, Phys. Lett.

B 516 (2001) 33.

\bibitem{tausk} J. B. Tausk, Phys. Lett. B 469 (1999) 225.

\bibitem{fleischer} J. Fleischer, V. A. Smirnov, A. Frink, J. K\"{o}rner, D.
Kreimer, K. Schilcher, J. B. Tausk, Eur. Phys. J. C2 (1998) 747.

\bibitem{hallyday} I. G. Halliday and R. M. Ricotta, Phys. Lett. B 193
(1987) 241; G. V. Dunne and I. G. Hallyday, Phys. Lett. B 193 (1987) 248.

\bibitem{suzuki/alexandre1} A. T. Suzuki, A. G. M. Schmidt, J. Phys. A:
Math. Gen. 31 (1998) 8023.

\bibitem{suzuki/alexandre2} A. T. Suzuki, A. G. M. Schmidt, Eur. Phys. J.
C12, (2000) 361.

\bibitem{suzuki/alexandre3} A. T. Suzuki, A. G. M. Schmidt, Eur. Phys. J. C
5 (1998) 175.

\bibitem{suzuki/alexandre4} A. T. Suzuki, A. G. M. Schmidt, Phys. Lett. B
494 (2000) 332.

\bibitem{davydy1} A. I. Davydychev and M. Yu. Kalmykov, (hep-th/0012189); A. I. Davydychev; A. G. Grozin,Eur. Phys. J. C 020 (2001) 333 (hep-ph/0103078); A. I. Davydychev, P. Osland, L. Saks, JHEP 0108 (2001) 050 (hep-ph/0105072).

\bibitem{simula} D.Melikhov, and S. Simula, (hep-ph/0112044)

\bibitem{hoof} G.\ `t Hooft and M. Veltman, Nucl. Phys. B 153 (1979) 365.

\bibitem{passarino} G. Passarino and M. Veltman, Nucl. Phys. B 160 (1979)
151.

\bibitem{brandt} F. T. Brandt and J. Frenkel, Phys. Rev. D 33 (1986) 464.

\bibitem{oldenborgh} G. J. Van Oldenborgh and J. A. M. Vermaseren, Z. Phys.
C 46 (1990) 425.
\end{thebibliography}
\end{document}